# POSITIONAL ACCURACY ASSESSMENT OF HISTORICAL GOOGLE EARTH IMAGERY

Peter C. Nwilo[1]., Chukwuma J. Okolie[1], Johanson C. Onyegbula[1], Ikenna D. Arungwa[2*], Owolabi Q. Ayoade[1,3], Olagoke E. Daramola[1], Michael J. Orji[1], Ikechukwu D. Maduako[4] and Imeime I. Uyo[1]

[1]Department of Surveying and Geoinformatics, Faculty of Engineering, University of Lagos, Nigeria
[2]Department of Surveying and Geoinformatics, Faculty of Engineering, Abia State University, Nigeria
[3]Bureau of Lands and Survey, Abeokuta, Ogun State, Nigeria
[4]Department of Geinformatics and Surveying, Faculty of Environmental Studies, University of Nigeria, Nsukka, Nigeria
*Corresponding author email: arungwaikenna@gmail.com

**ABSTRACT**

Google Earth (GE) is the most popular virtual globe in use today. Given its popularity and usefulness, most users do not pay close attention to the positional accuracy of the imagery, and there is limited information on the subject. This study therefore evaluates the horizontal accuracy of historical GE imagery at four epochs between year 2000 and 2018, and the vertical accuracy of its elevation data within Lagos State in Nigeria, West Africa. The horizontal accuracies of the images were evaluated by comparison with a very high resolution (VHR) digital orthophoto while the vertical accuracy was assessed by comparison with a network of 558 ground control points. The GE elevations were also compared to elevation data from two readily available 30-metre digital elevation models (DEMs) – the Shuttle Radar Topography Mission (SRTM) v3.0 and the Advanced Land Observing Satellite World 3D (AW3D) DEM v2.1. In terms of the horizontal accuracy, the root mean square errors (RMSEs) are as follows – year 2000 (29.369m), year 2008 (28.391m), year 2012 (10.615m) and year 2018 (10.603m). The most recent GE imagery (year 2018) was the most accurate while year 2000 was the least accurate. This shows a continuous enhancement in the accuracy and reliability of satellite imagery data sources which form the source of Google Earth data. Results also portray that the GE images have a tendency to be skewed towards the western and north-western  directions, indicative of systematic error. In terms of the vertical accuracy, GE elevation data had the highest RMSE of 6.213m followed by AW3D with an RMSE of 4.388m and SRTM with an RMSE of 3.682m. Although the vertical accuracy of SRTM and AW3D are superior, Google Earth still presents clear advantages in terms of its ease-of-use and contextual awareness.

## 1.    INTRODUCTION

The integration of spatial technologies with the world wide web has led to the evolution of virtual globes which provide worldwide access to geospatial data (Elvidge and Tuttle, 2008). Allen (2008) defines a virtual globe as "a 3D software model of the Earth (or other planet) that provides a user interactivity and freedom to view the globe from different viewing angles, positions, and overlays of actual or abstract geographic data." The ease of use of digital virtual globes and their capacity for display and visualisation of spatial information make them a powerful communication tool for researchers, decision makers and the general public (Aurambout et al., 2008). Virtual globes present a simpler alternative to technocratic and costly Geographic Information System (GIS) software, and this facilitates sharing of spatial data at a global scale (Yu and Gong, 2012). Virtual globes can be viewed as technological realisations of the Digital Earth (DE) concept introduced by former United States Vice President Al Gore (Gore, 1998; Liang et al., 2018); and have led to new paradigms in the concept of Digital Earth (Goodchild et al., 2012; Pulighe et al., 2016). Digital Earth has been described as "a multiresolution and three-dimensional visual representation of Earth that would help humankind take advantage of geo-referenced information on physical and social environments, linked to an interconnected web of digital libraries" (Gore, 1999 in Liu et al., 2020).

Examples of free and publicly available virtual globes/image services include Google Earth, Google Maps, NASA World Wind, Microsoft Bing Maps and Apple Maps (Pulighe et al., 2016; Goudarzi and Landry, 2017). Among these examples, Google Earth (GE) is the most popular and versatile. It renders a three-dimensional (3D) representation of Earth by the superimposition of images obtained from satellite imagery with worldwide coverage, aerial photography from local or national mapping agencies, near-orthophoto collections in GeoPortals and GIS 3D globe. Google Earth can show various kinds of images overlaid on the surface of the earth and is also a Web Map Service client. The core technology behind Google Earth was originally developed at Intrinsic Graphics in the late 1990s. In version 5.0, Google introduced "Historical Imagery", allowing users to view images of a region at different epochs and to observe an area's changes over time (see Figure 1). 3D coverage of cities by Google Earth began in 2012 (Ubukawa, 2013). By early 2016, it had been expanded from 21 cities in 4 countries to hundreds of cities in over 40 countries, including every US state and encompassing every continent except Antarctica. The very high resolution (VHR) satellite images on Google Earth have a spatial resolution finer than 5m (Lesiv et al., 2018). However, the spatial resolution of the images depends on the characteristics of the satellite such as the altitude and type of instruments (Buka et al., 2015). In reality, GE images are not spatiotemporally continuous or homogenous but are mosaicked using multiple images from different periods, different spatial resolutions ranging from 15m to 10cm, and from different imagery providers (Lesiv et al., 2018). The images are compiled from a wide variety of sources such as: SPOT 5, Rapid Eye, Earth Resource Observation Satellites (EROS), Meteosat 2, Geoeye 1, and Digital Globe World View 2 satellite (Buka et al., 2015). Since Google Earth images are sourced from multiple sources, they do not have identical positional accuracy or spatial resolution (Goudarzi and Landry, 2017). The satellite images are sometimes supplemented with aerial photographs which have a higher resolution. In places where high resolution imagery is unavailable, GE defaults to Landsat imagery (Potere, 2008). On the frequency of updates, Google aims to update satellite imagery of places that undergo frequent changes, once a year for big cities, every two years for medium-sized cities and up to every three years for smaller cities (Schottenfels, 2020).

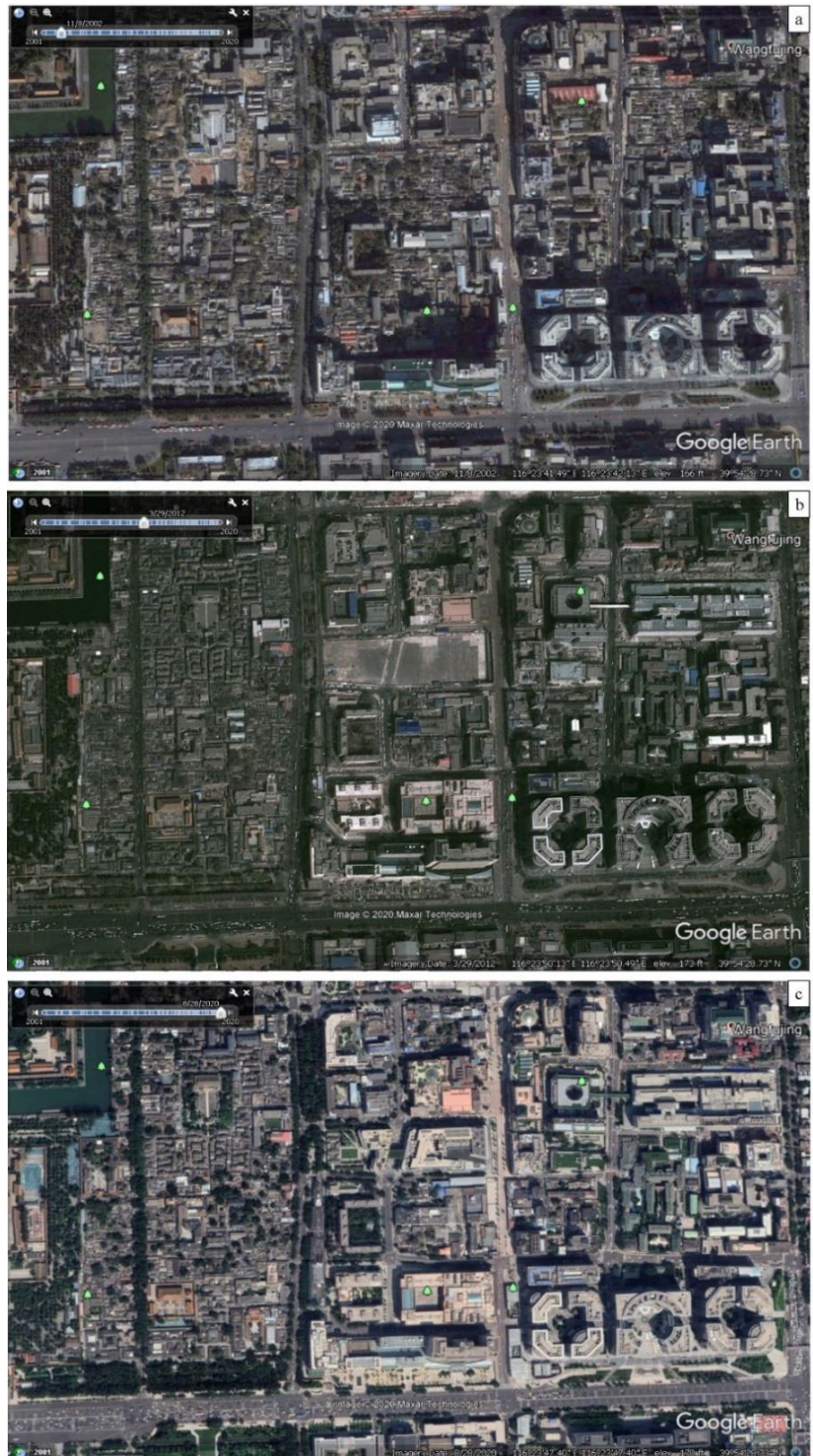

Figure 1: Historical Google Earth imagery over a part of Beijing city in China at three periods – (a) 8 November, 2002 (b) 29 March, 2012 (c) 28 August, 2020. The Historical Imagery slider is visible at the top left corner of the images

There is some ambiguity on the source of Google Earth elevation data (Goudarzi and Landry, 2017). It is possibly derived from the Shuttle Radar Topography Mission (SRTM) DEM, the Advanced Spaceborne Thermal Emission and Reflection Radiometer (ASTER) Global DEM, and from Light Detection and Ranging (LiDAR) (Goudarzi and Landry, 2017, Chigbu et al., 2019; MES Innovation Sdn Bhd, 2020). The recent introduction of elevation data sourced from LiDAR, interestingly makes it possible for half accuracy of about 5-25cm. Ironically, the location of such places where LiDAR data covers are not known or revealed by GE (MES Innovation Sdn Bhd, 2020). It is expected that errors inherent in the elevation data sources would naturally propagate into GE elevation data.

Since the launch of Google Earth in 2005, it has enjoyed ever-increasing popularity from map makers, pathfinders, navigators, planners, application developers, etc as a free data source providing a realistic view of the world through satellite images, maps, digital terrain, 3D buildings, land use information, identification of monuments and locational data. Google Earth imagery has found wide applications in health geography research (Curtis et al., 2006), land use/land cover mapping (Hu et al., 2013; Malarvizhi et al., 2016), land conversion studies (Jacobson et al., 2015), mapping of lakes (Shen et al., 2006), internet GIS (Henry, 2009), urban household surveys (Ngom Vougat et al., 2019), real estate (Hwang, 2008), and relief/humanitarian efforts (Nourbakhsh et al., 2006). GE Historical Imagery provides images taken at different periods and this has wide applications in land use change detection studies (Malarvizhi et al., 2016). Generally, the use of GE in research projects have been summarised into the following categories: visualisation, data collection, validation, data integration, communication/dissemination of research results, modelling, data exploration and decision support (Yu and Gong, 2012). In the scientific community, its use pertains to earth surface processes, habitat availability, health and surveillance systems, biology, land use/land cover (LULC), agriculture, landscape etc. (Pulighe et al., 2016). Comprehensive reviews of earth science applications of Google Earth are provided in Yu and Gong (2012) and Liang et al. (2018).

Google Earth presents a new paradigm in Digital Earth and in the quest by man to understand the environment and effectively manage its resources. It also presents a clear advantage to achieving the United Nations 2030 Agenda for sustainable development. As a virtual globe, Google Earth connects all parts of the world in a virtual environment with free access to geospatial data to support global partnerships in attaining the sustainable development goals (SDGs). More so, policy and decision making at every level (local, national, regional or global) are dependent on up-to-date geospatial data. Globally, there is a continuous drive by policy makers to deliver sustainable development within, and in accordance with the templates provided by the SDGs. The SDGs are earth-centred and driven by geospatial data. For example, without geospatial data in place, the idea of location-based services would to a large extent remain a mirage. Google Earth is therefore relevant for achieving SDG 2 (zero hunger), SDG 6 (clean water and sanitation), SDG 9 (industry, innovation and infrastructure), SDG 11 (sustainable cities and communities), SDG 13 (climate action), SDG 14 (life below water), SDG 15 (life on land) and SDG 7 (affordable and clean energy). As a virtual globe, Google Earth is also a crucial tool for bridging the global North – South divide in terms of access to geospatial data for international partnerships and collaborations.

Given the popularity of Google Earth, users tend to assume that it is a highly accurate source of information with no doubt on its positional accuracy (Flanagin and Metzger, 2008). However, there are questions surrounding the reliability of GE imagery, since very little is known about its metadata including the sensors, imagery resolutions, and overlay/mosaicking techniques (Pulighe et al., 2016). According to Paredes-Hernández et al. (2013), Google geographic data products are only approximations without officially documented accuracies. Wang et al. (2017) note that Google has been unwilling to release comprehensive information on the accuracy of the GE archive. It is also mentioned that GE images are also not orthorectified and lack photogrammetric accuracy (Goudarzo and Landry, 2017). The uncertainty surrounding the horizontal accuracy of the imagery could lead to feature misrepresentations, and incorrect inferences (McRoberts 2010 in Pulighe et al., 2016). There are also errors in image alignment manifesting at the transition zones between mosaicked images on Google Earth (Potere, 2008) (e.g., disjoint shorelines and roads shown in Figure 2). This presents some uncertainty on the usability of GE imagery for sensitive applications requiring very high accuracy such as high-precision engineering surveys and autonomous navigation. The practice of reporting coordinates with a precision that does not match its accuracy misleads users to believe that it is an accurate source of information (Goodchild et al., 2012). Moreover, Benker et al. (2011) noted that Google's representatives stated that the coordinates provided by Google and the data available in their geographic products are only approximations and that Google makes no claim to the accuracy of their geographic information products. A quick check of the GE historical images at some locations (Figure 3) shows that the magnitude of these horizontal shifts varies with time. In some cases, the positional errors are not consistent when viewed at different periods with the Historical Imagery slider (see Figure 3). Another limitation is that little is known about the volume of historical imagery in Google Earth's archive and where it can be found (Lesiv et al., 2018).

Positional accuracy is traditionally divided into two classes: horizontal accuracy and vertical accuracy (Goudarzi and Landry, 2017). Becek et al. (2011) identified the flaws associated with the positional identities of some known points from Global Elevation Data Testing Facility (GEDTF) and their corresponding points on GE imagery. A remarkable error of more than 1.5km was noticed in some cases after measuring the discrepancies using some tools and basic statistics. In Paredes-Hernández et al (2013), geo-registration and large horizontal errors were shown to occur in GE imagery. However, the authors suggested the possibility of GE imagery satisfying the horizontal accuracy requirements of the American Society for Photogrammetry and Remote Sensing (ASPRS), assessed in terms of root mean square error (RMSE) for x, y and z coordinates, for the production of "Class 1" 1: 20,000 maps, if a large number of well-defined points are extracted from areas of high-resolution imagery over rural areas. Mulu and Derib (2019) evaluated the accuracy of GE imagery in Khartoum, Sudan and showed that the horizontal RMSE was suitable for producing a Class 1 map of 1:20,000 scale (as recommended by ASPRS, 1990). However, they pointed out that the resolution of the acquired Google Earth imagery was a major factor affecting the accuracy of the GE dataset, as coarser resolutions appeared to have higher RMSE values probably due to less accurate location of control points on such coarse resolutions. Goudarzi and Landry (2017) assessed the horizontal accuracy of GE in the city of Montreal, Canada using ten Global Positioning System (GPS) reference points. In their results, the positional accuracy varied between ~0.1m in the south to ~2.7 m in the north of the city.

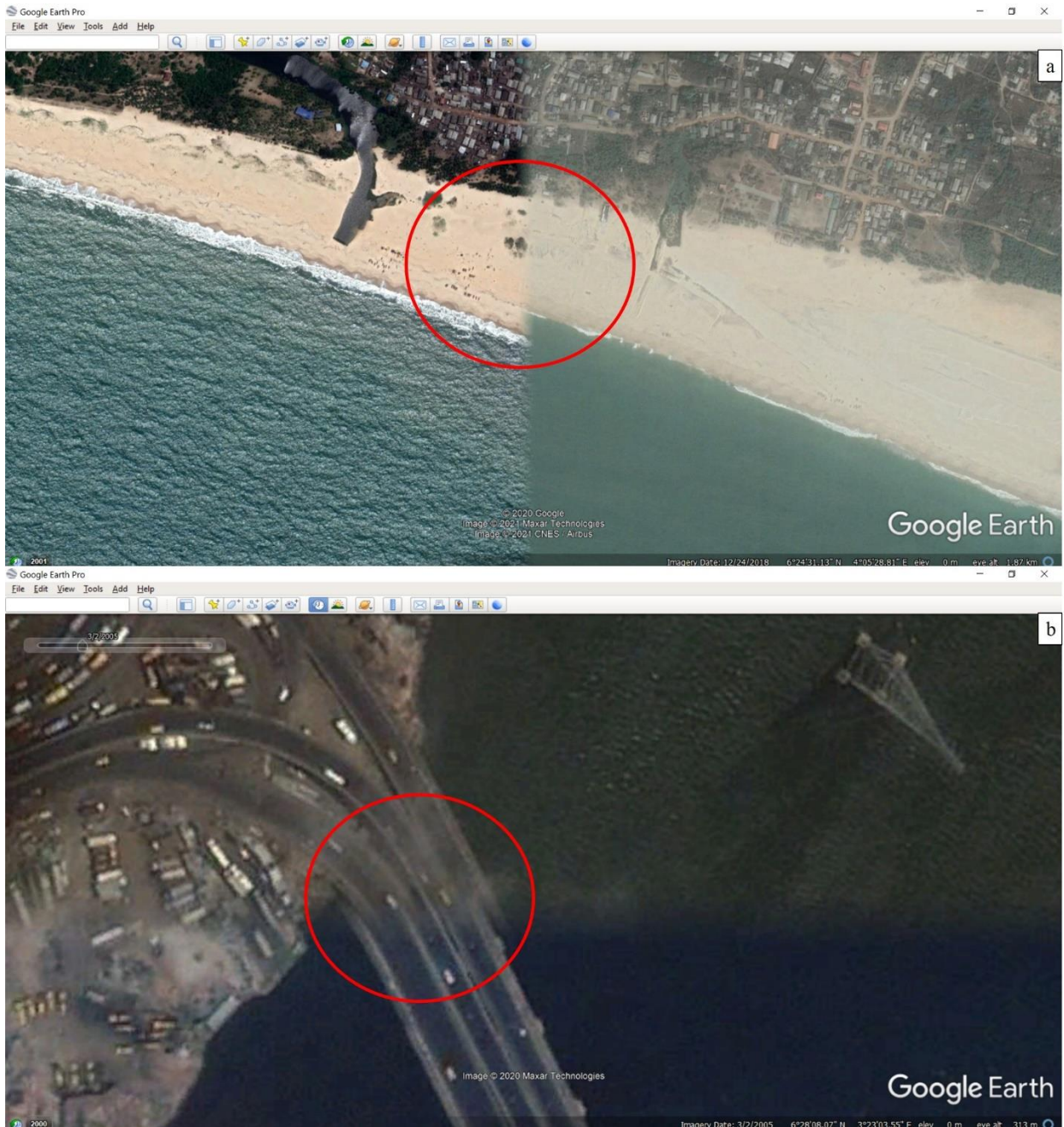

Figure 2: Region of image misalignment (within the red circles) on Google Earth imagery in Lagos, Nigeria - (a) disjoint shoreline, imagery date - December 2018 (b) disjoint roads along Carter Bridge, imagery date – March 2005

According to El-Ashmawy (2016), the accuracies of DEMs prepared from GE data are only suitable for certain engineering applications but inadequate for very precise engineering studies. It might satisfy the vertical accuracy requirements of the ASPRS (1993) standards for the production of "Class III" contour maps. Other applications of GE elevation data include the preparation of large-area cadastral, city planning, or land classification maps. In Aba metropolis of south-eastern Nigeria, Chigbu et al (2019) assessed GE elevation data using a 10.16km elevation profile data obtained by means of ground survey as reference. They reported a mean error of 1.65m, RMSE of 2.79m, standard deviation of 2.27m and median absolute deviation of 1.72m for the GE elevations. However, on the strength of further incisive statistical tests (Mann-Whitney U Test of group and the t-Test), they concluded that GE elevation data was unfit for any form of levelling operation that would eventually lead to engineering construction.

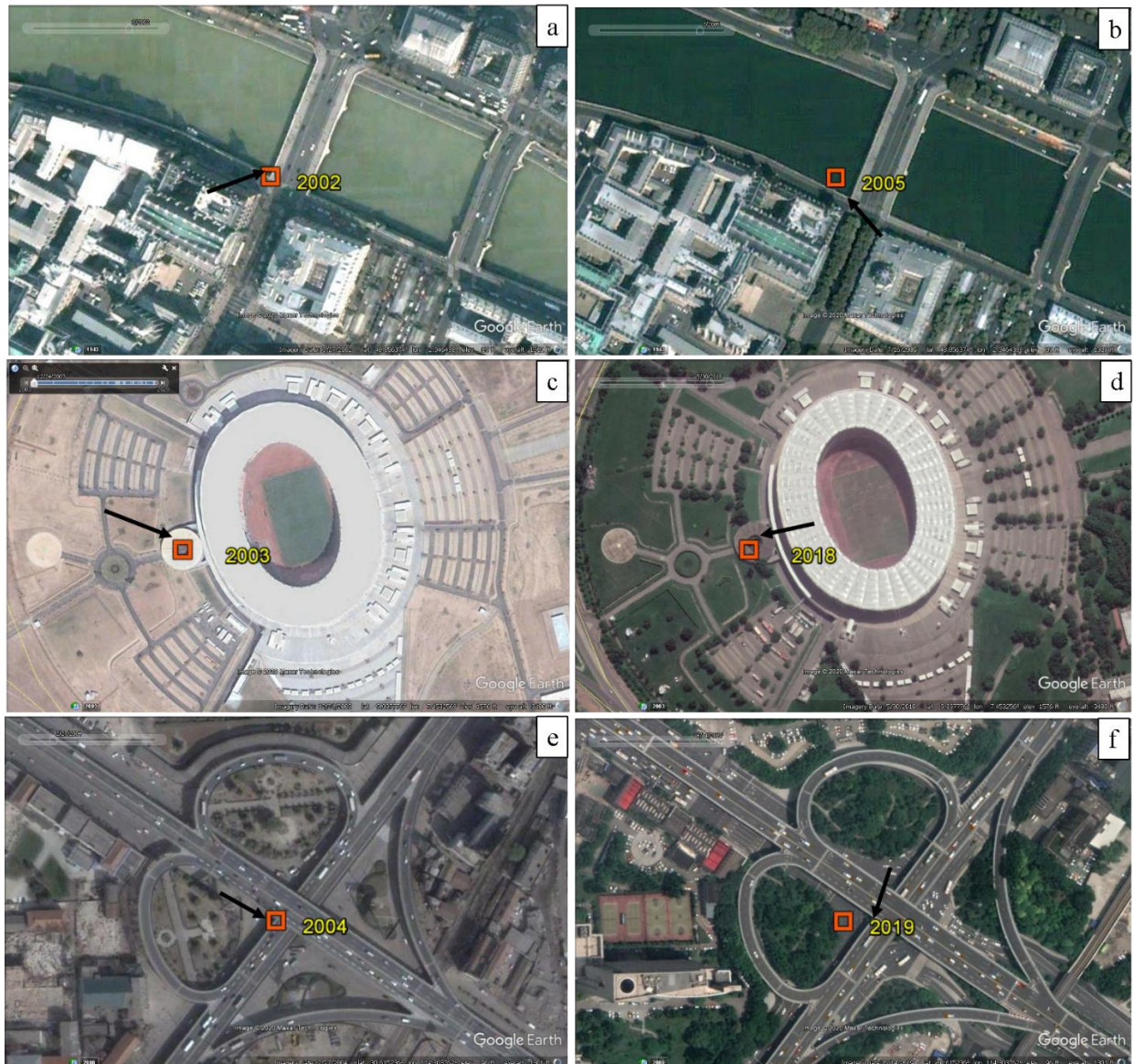

Figure 3: Horizontal shifts in positions of features on Google Earth shown at three locations - Pont au Change bridge on the Seine River in Paris France (a) 2002 (b) 2005; National Stadium in Abuja Nigeria (c) 2003 (d) 2018; Yonghongcun in Wuhan China (e) 2004 (f) 2019

The issue of GE's positional accuracy has received little interest from researchers around the world. Most of the studies discovered in the literature survey, focused only on the horizontal accuracy and there was little interest in the vertical accuracy of its elevation data. Moreover, a literature search did not reveal any studies dealing with the issue of horizontal error in historical GE imagery. Errors in the geo-registration of GE images could limit the scientific value of the archive (Potere, 2008). Hence, the present study investigated the horizontal accuracy of historical

GE imagery at four periods between 2000 and 2018, and the vertical accuracy of its elevation data within Lagos State in Nigeria, West Africa. The horizontal accuracies of the images were assessed by comparison with a highly accurate digital orthophoto while the vertical accuracy was assessed by comparison with a network of ground control points. The GE elevations were also compared to elevation data from two publicly available 30-metre DEMs – the Shuttle Radar Topography Mission (SRTM) v3.0 DEM and the Advanced Land Observing Satellite World 3D (AW3D) DEM v2.1. To our knowledge, this is the first study to assess the horizontal accuracy of Google Earth "historical imagery". The findings are important to inform users of the reliability of GE imagery for use in change detection studies and other analyses that involve spatio-temporal variability. It also provides a critical knowledge base to inform end-users on the quality and reliability of the data for a myriad of applications.

## 2.    MATERIALS AND METHODS

The study involved acquisition of various datasets relevant to the evaluation of the positional accuracy. The steps involved in realising this are as outlined in sub-sections below.

### 2.1    Study Area

The study area is Lagos State in Nigeria, West Africa. Lagos State is located in the south-western part of Nigeria (Figure 4) and is bounded in the south by the Atlantic Ocean, in the west by the Republic of Benin and in the north and east by Ogun State. The state was once the administrative capital of Nigeria between 1976 and 1991. It is currently the commercial capital, a beehive of commercial and industrial activities, and has an estimated population of over 24 million (Lagos Digest of Statistics, 2017). It has a total area of about 3,577.28km$^2$, of which 2,792.72km$^2$ is covered by land and 779.56km$^2$ is water (BudgIT, 2018). The state is geographically located between longitudes 2º41'55'E – 4º22'00'E and latitudes 6º22'22"N – 6º43'20"N. It has a generally low-lying terrain with the Lagos and Lekki lagoons as its major water bodies. There are two major climatic seasons: the rainy season and the dry season. The mangrove swamp forest and freshwater swamp constitute some of its most dominant vegetation types. Temperature ranges from 20ºC - 32ºC and the mean annual rainfall exceeds 1,700mm (Nwilo et al., 2020). For the horizontal accuracy assessment in this study,  a digital orthophoto of the University of Lagos was acquired. The University of Lagos is one of the federal universities in Nigeria, situated within metropolitan Lagos. It is located between longitudes 3º23'00"E – 3º24'30"E and latitudes 6º30'00"N – 6º31'30"N. As an institution for learning and research, it is surrounded by research infrastructure, buildings and commercial activities, and also bounded eastwards by the Lagos Lagoon.

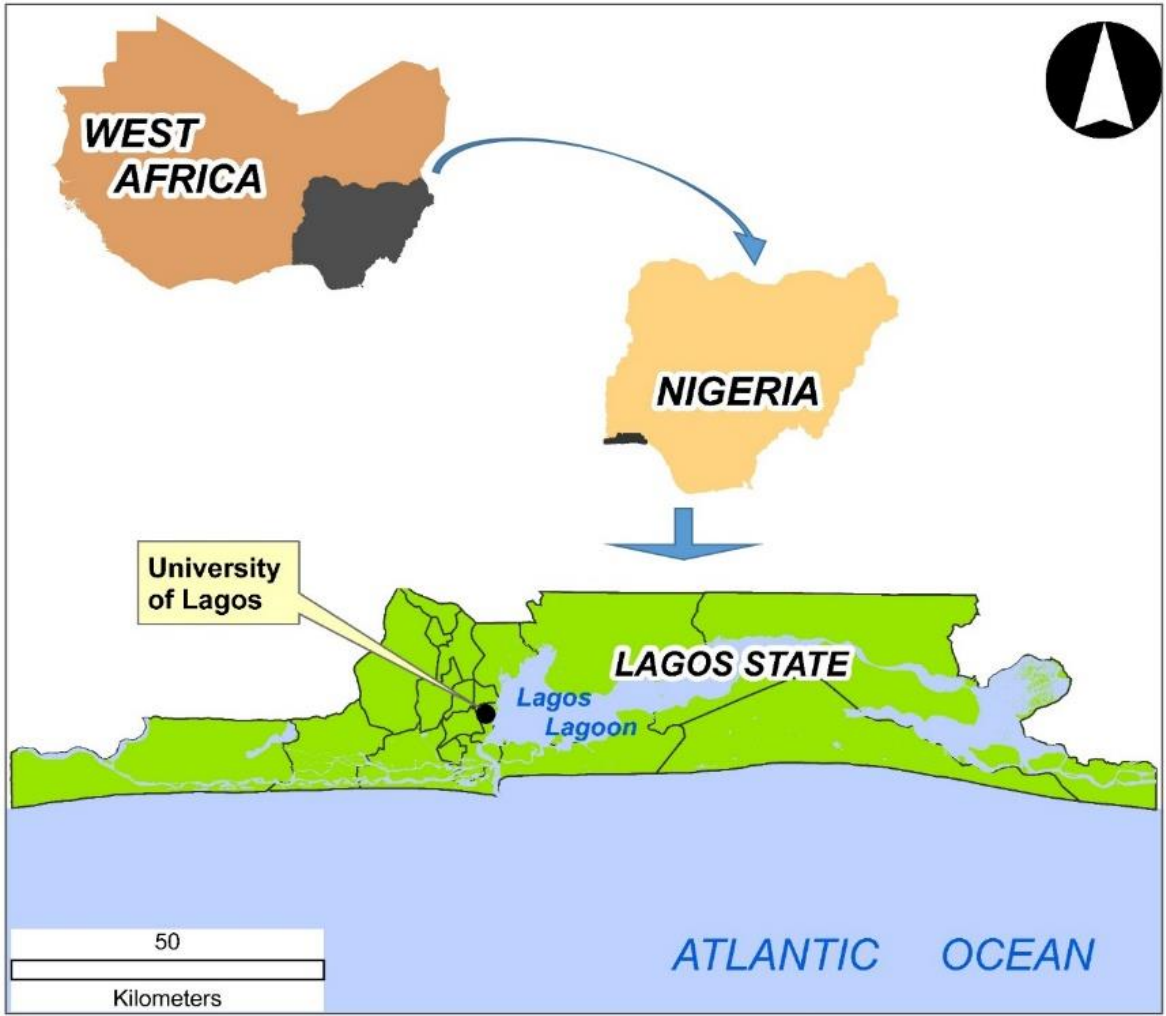

Figure 4: Map showing location of the study area

## 2.2 Description of Datasets

The datasets used are discussed below.

### 2.2.1 Ground control points

The rectification of the orthophotos was done using highly accurate ground control points (GCPs) within the University of Lagos. The GCPs were surveyed with the Trimble R8 dual frequency Global Navigation Satellite System (GNSS). The GNSS field procedure and data processing is well explained in Okolie (2019) and Gbopa et al. (2021). Essentially, fourteen GCPs (shown in Figure 5) were signalised on the ground. Ground control signalisation is the selection of ground control identification style or pattern. The signalisation was done with cross markings using white emulsion paint to ensure their visibility from a high altitude during the Unmanned Aerial Vehicle (UAV) survey. The cross markings were approximately 80 – 100cm in length and 15 – 20cm in breadth. Figure 6 shows two of the signalised GCPs within the university, YTT 28/186 and XST 347. The GNSS observation was then carried out on the GCPs in static mode with about 30 – 40 minutes occupation time on each point. After completion of the survey, the data was downloaded from the GNSS receivers and post-processed to derive the final coordinates. For the vertical accuracy assessment, 558 geodetic GCPs in Lagos State were acquired from the Lagos State Surveyor General's Office.

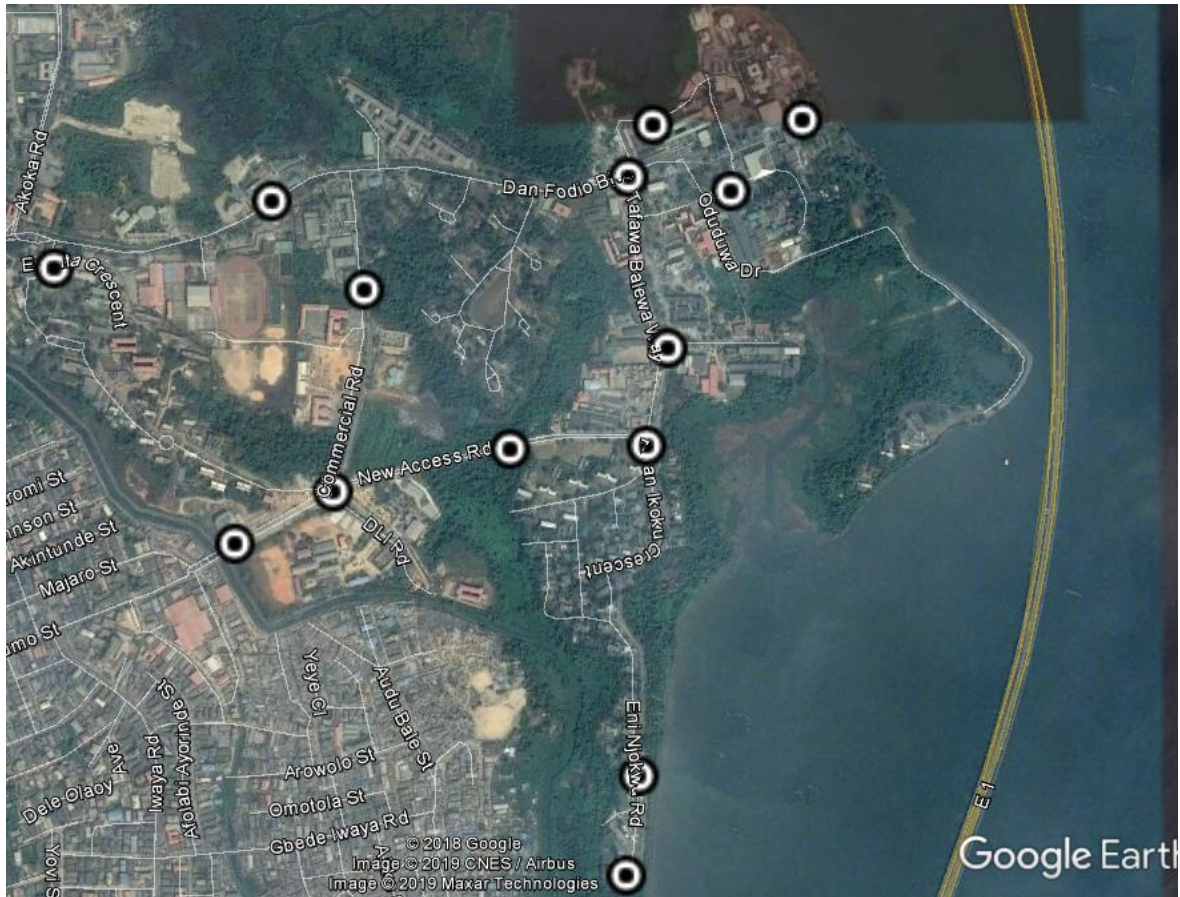

Figure 5: Spatial distribution of the signalised GCPs

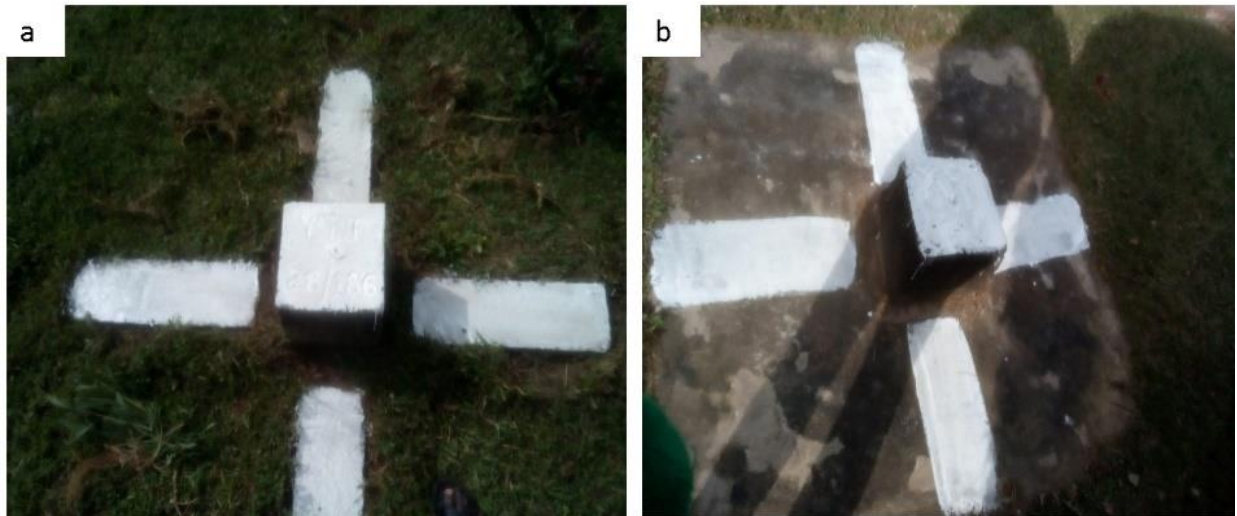

Figure 6: Some of the signalised GCPs within the University of Lagos – (a) YTT 28/186 and (b) XST 347

### 2.2.2  Digital orthophoto

The reference dataset for the horizontal accuracy assessment was a digital orthophoto which was derived from overlapping images of the University of Lagos main campus captured during a UAV survey conducted in July 2019. A detailed discussion of the UAV flight and orthophoto processing is discussed in Okolie (2019) and Gbopa et al. (2021). The DJI Phantom 4 Professional UAV with a camera lens of 84° field of view (FOV) and a focal length of 8.8 mm/24 mm (35 mm format equivalent) was used for the survey. The gimbal has a controllable range of −90° to +30° and a maximum controllable angular speed of 90°/s (Gbopa et al., 2021). The flight was conducted at an altitude of 90m, a UAV speed limit of 15m/s, flight direction of 126°, and with generous overlaps of 75% fore and 65% side. Figure 7 shows the UAV flight in progress at two locations within the University of Lagos  while Figure 8 shows some images captured from the UAV survey. After the flight, the raw photos were downloaded and imported into Pix4D Mapper software environment. The processing with Pix4D Mapper addressed the image alignment, geo-rectification with GCPs, the processing of point cloud and mesh, and orthophoto generation. The orthophoto was produced at a ground sampling distance (GSD) of 4.36cm. Gbopa et al. (2021) evaluated the accuracy (RMSE) of the orthophoto using 7 GCPs of high fidelity and obtained a horizontal and vertical accuracy of 0.183m and 0.157m respectively. This shows a high level of accuracy adequate for use as a reference dataset to evaluate the horizontal accuracy of the GE images. Figure 9 shows the final orthophoto map with 45 points initially selected as reference points for the horizontal accuracy assessment.

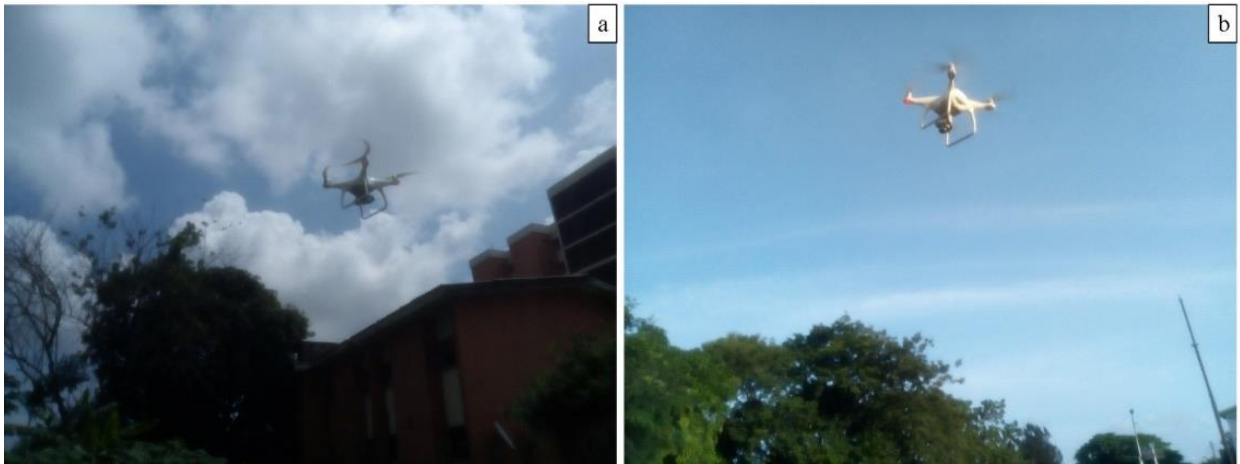

Figure 7: UAV flight in progress within the University of Lagos main campus

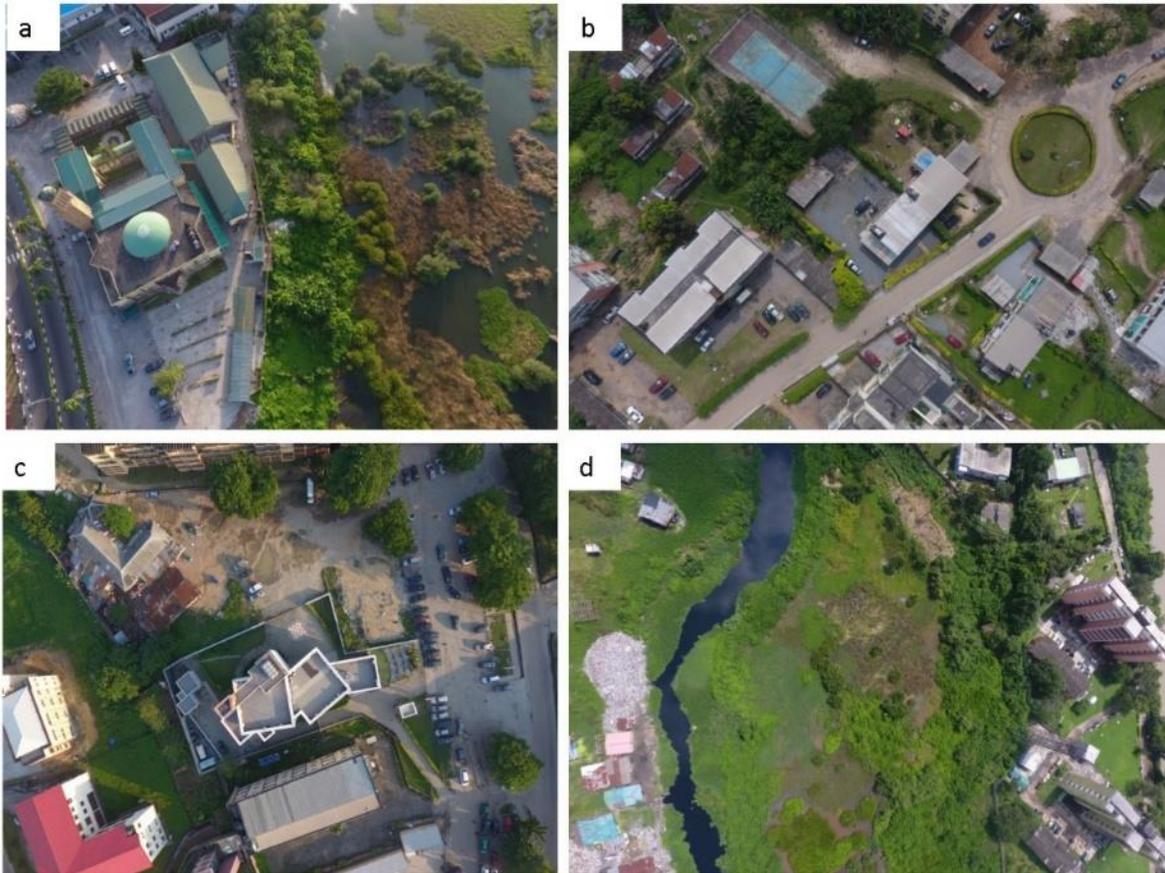

Figure 8: Some images from the UAV survey (a) Wetlands to the right of the University of Lagos Central Mosque (b) Staff Quarters (c) GTBank building (centre) and Erastus Akingbola Hall (bottom centre) (d) High-Rise Complex (far-right) and a canal running along the southern boundary of the campus

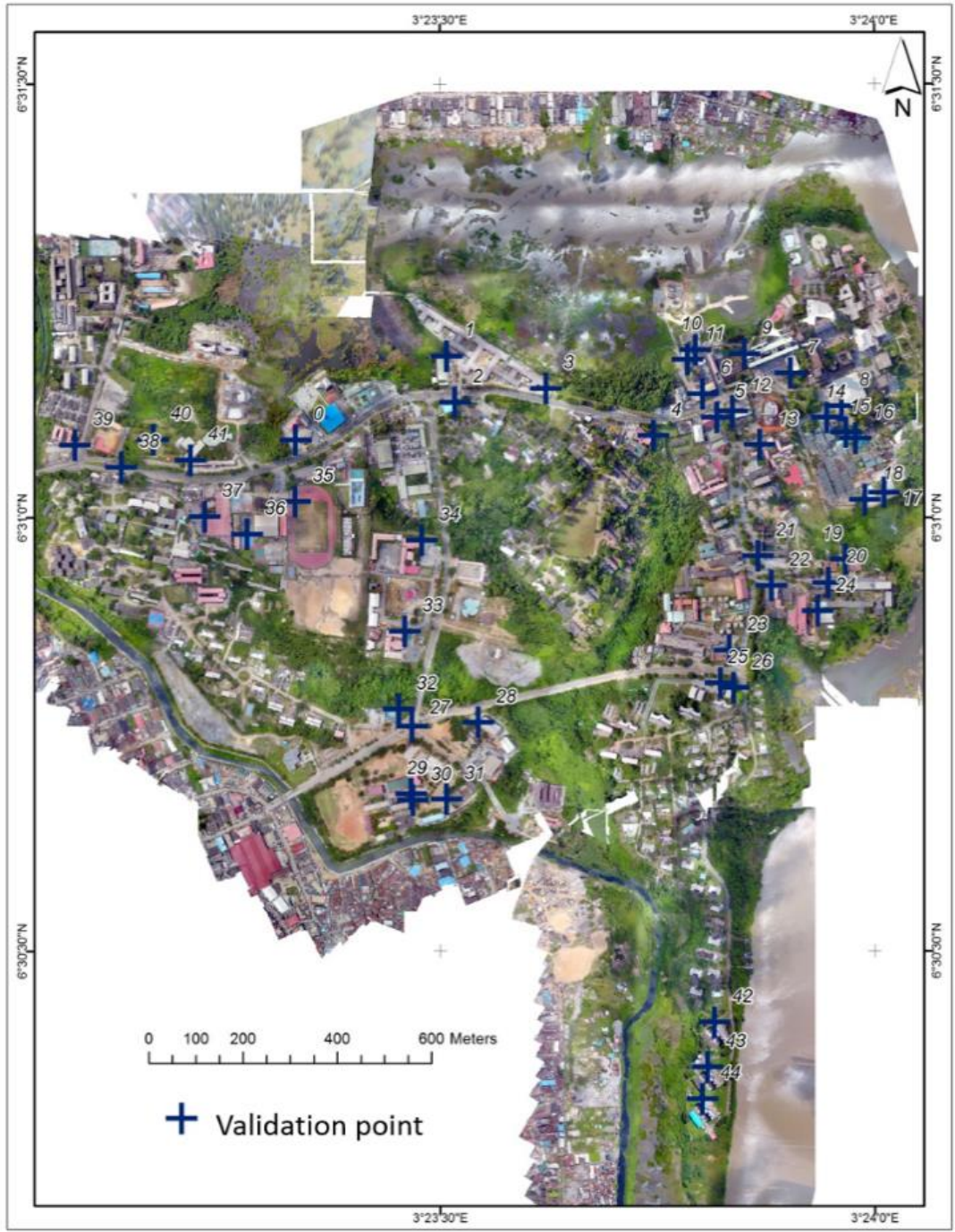

Figure 9: Orthophoto map of University of Lagos showing the spatial distribution of the initial 45 reference points selected for horizontal accuracy assessment (27 points were retained in the final selection)

### 2.2.3   Google earth historical imagery

The Historical Imagery tool in Google Earth Pro v.7.3 enables the viewing of images at different epochs, and allows for observation of an area's changes over time. Historical Google Earth images between the years 2000 and 2018 were considered. The criteria for selecting a particular year included the image clarity and degree of cloud cover. After a meticulous inspection, the following periods were selected: 13th December 2000, 7th October 2008, 4th June 2012 and 28th December 2018. Google Earth images are referenced to the WGS84 datum which is based on the GRS80 ellipsoid (Buka et al., 2015). Figure 10 shows the points selected on the historical images to coincide with the reference points from the orthophoto.

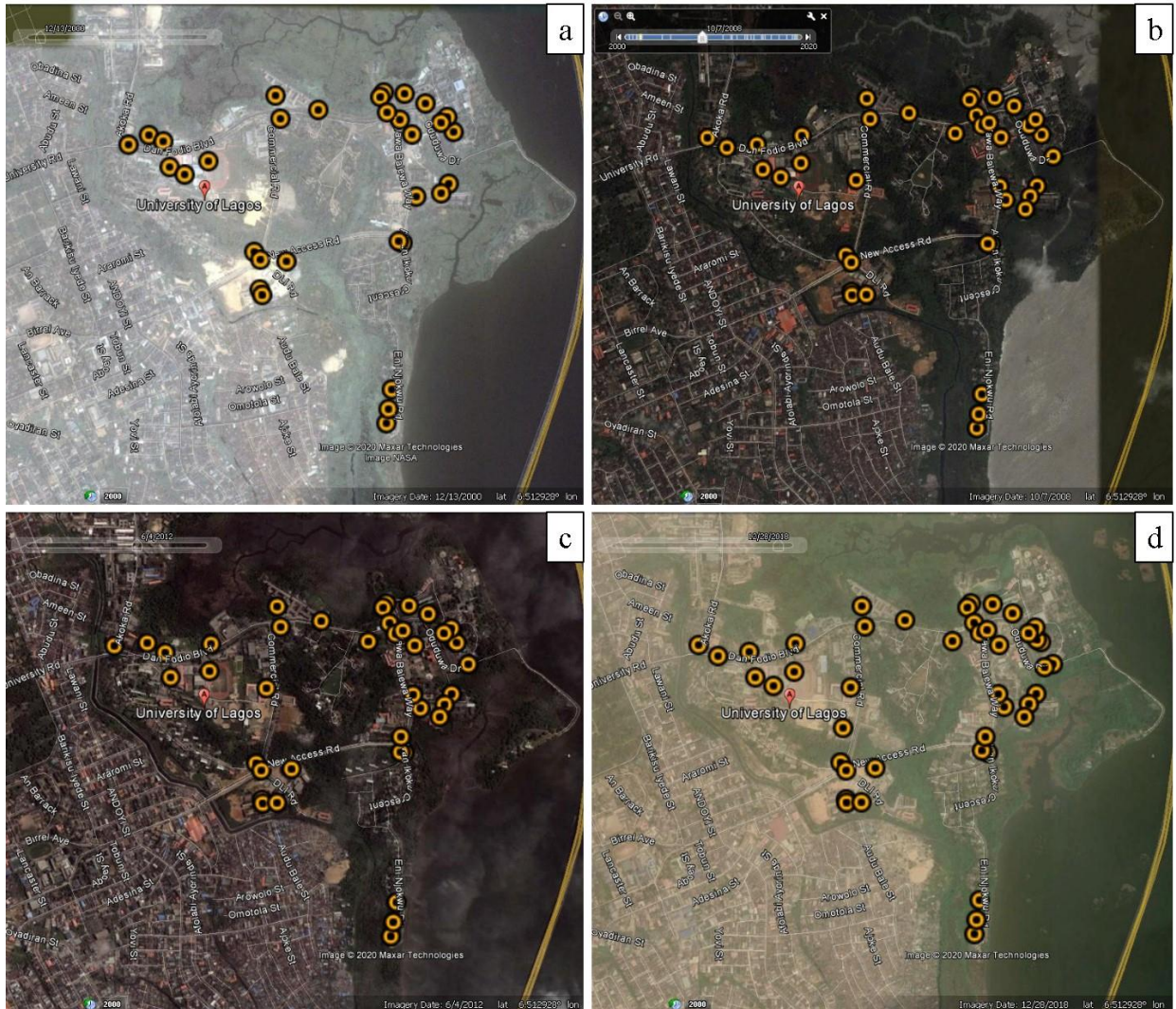

Figure 10: Coincident points on Google Earth imagery for comparison with the reference pints on the orthophoto (a) 13th December, 2000 (b) 7th October, 2008 (c) 4th June 2012 (d) 28th December, 2018

## 2.2.4   Google earth elevation data, SRTM v3.0 and AW3D30 v2.1

Elevation data were extracted from Google Earth at points coincident with the 558 GCPs in Lagos State. The SRTM v3.0 DEM covering Lagos State was downloaded from the United States Geological Survey (USGS) Earth Explorer website (https://earthexplorer.usgs.gov/). The SRTM project was jointly executed by the National Aeronautics and Space Administration (NASA) and the National Imagery and Mapping Agency (NIMA) of the US Department of Defense. The mission was designed to use a single-pass radar interferometer to produce a digital elevation model (DEM) of the Earth's land surface between $60^0$N and $56^0$S latitude (Farr and Kobrick, 2000). The SRTM DEM of 1 arc-second resolution which corresponds to 30m spatial resolution at the equator is distributed in 1° × 1° tiles, and has higher accuracy than the earlier 90m SRTM DEM product (Mukul et al., 2017). The SRTM mission goal of LE90 error of 16m (RMSE ~10 m) was assessed worldwide and validated using dual-frequency, Real Time Kinematic (RTK) GPS data (Üstün et al., 2016). In the published global assessment report of SRTM DEM, it is stated that the vertical accuracy meets and exceeds the performance requirements of the mission by a factor of nearly two in comparison to ground-truth data such as kinematic GPS trajectories on road networks (Mukul et al., 2015). The EGM96 geoid provides the vertical datum for SRTM.

In 2015, the Advanced Land Observing Satellite (ALOS) World 3D - 30m (AW3D30) DEM was made available as an Earth topography elevation data product (https://www.eorc.jaxa.jp/ALOS/en/aw3d30/index.htm). AW3D30 was photogrammetrically developed using optical imagery collected during the ALOS mission (Tadono et al., 2016). The ALOS elevation maps were produced at a spatial resolution of 5m with an accuracy of 5m (standard deviation). In 2016, JAXA released a 30m product, the AW3D30 dataset which was generated from the earlier 5m product (JAXA 2017). From the accuracy assessments by Caglar et al. (2018), the AW3D30 surpassed the accuracies of SRTM and ASTER GDEM Version 2. Figure 11 presents the spatial distribution of the 558 GCPs used for vertical accuracy assessment with SRTM v3.0 and AW3D30 v2.1 shown as the backdrop.

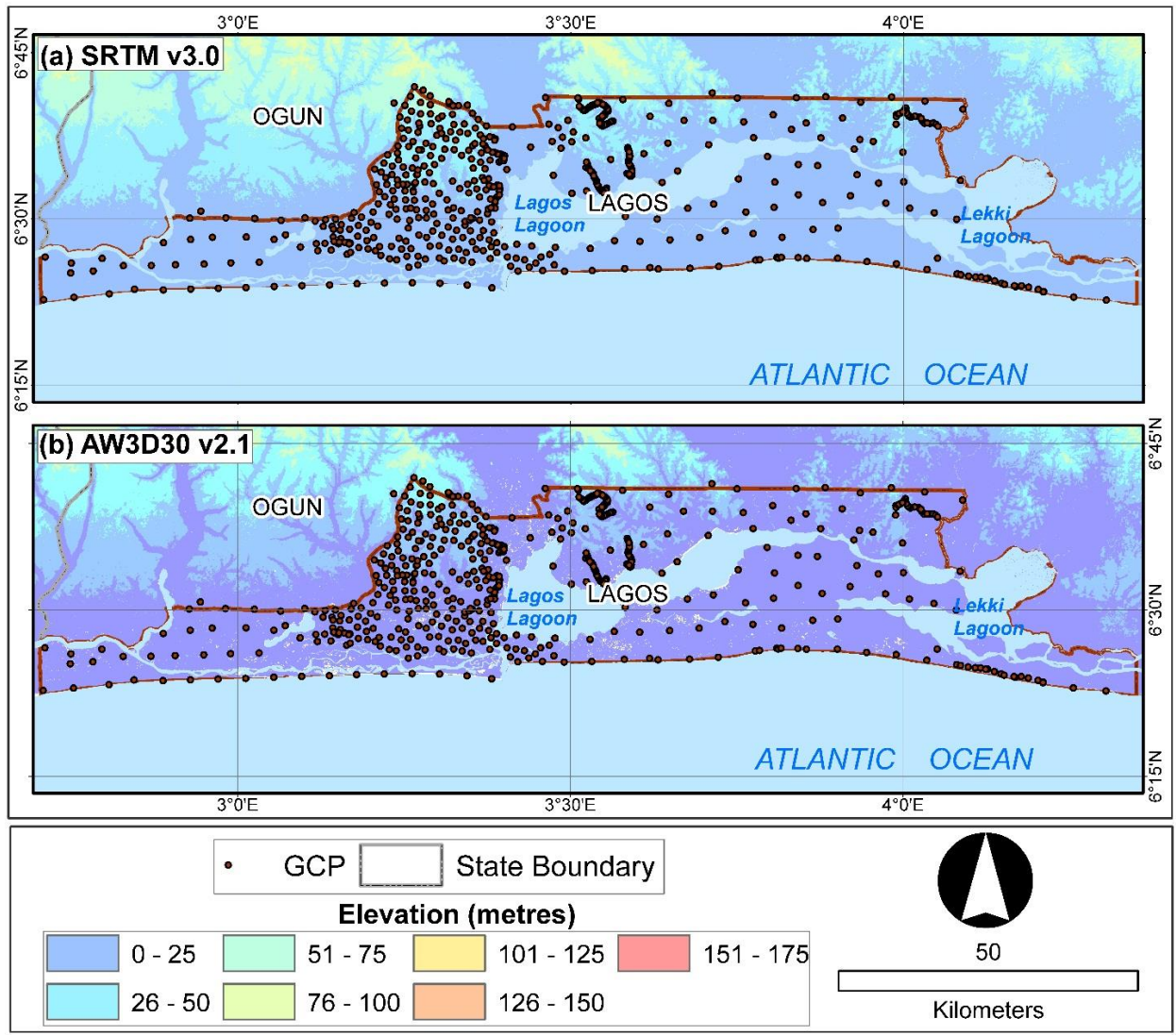

Figure 11: Overlay of 558 geodetic GCPs for vertical accuracy assessment on elevation map of (a) SRTMv3.0 and (b) AW3D30 v2.1

## 2.3    Data Exploration and Extraction

Forty-five points (#0 to #44) were initially marked as reference points on the orthophoto for use in the horizontal accuracy assessment. The points were carefully placed in easily identifiable locations such as road intersections, roundabouts, and corner points of buildings. The dates were set on Google Earth using the historical imagery time slider and the easting and northing coordinates of corresponding points on Google Earth were extracted. After inspecting the positions of all the points on the orthophoto and Google Earth images, it was observed that some points on the Google Earth images were located in problematic places such as cloud-covered areas etc. The problematic points were eliminated thus reducing the final selection to 27 points. To extract the Google Earth elevations for vertical accuracy assessment, the 558 GCPs were first converted to keyhole markup language (kml/kmz) format and imported into Google Earth. Next,  each point was zoomed to and the corresponding elevation on Google Earth was copied from the status bar and pasted directly in a Microsoft Excel sheet.   The corresponding elevations from SRTM and AW3D were extracted in the ArcGIS 10.1 environment using the 'extract values to points' tool in Spatial Analyst.

## 2.4    Quantitative Analysis

The coordinate differences between the orthophoto reference points and the coincident points on the Google Earth images were used as a basic metric in assessing the horizontal accuracy. Coordinates of points on the orthophoto were subtracted from those of the various epochs for the GE historical imagery. The differences in eastings and northings are given as follows:

$$\Delta E = E_{GE} - E_{OP} \tag{1}$$
$$\Delta N = N_{GE} - N_{OP} \tag{2}$$

*Where,*

$\Delta E$ and $\Delta N$ are the differences in Eastings and Northings respectively for the selected points. $E_{GE}$ and $E_{OP}$ are Eastings of selected points on the GE historical imagery and orthophoto respectively. $N_{GE}$ and $N_{OP}$ are Northings of selected points on the GE historical imagery and orthophoto respectively. The linear separation between the position of a point on the orthophoto and on Google Earth is the horizontal error/shift (S) (Figure 12).

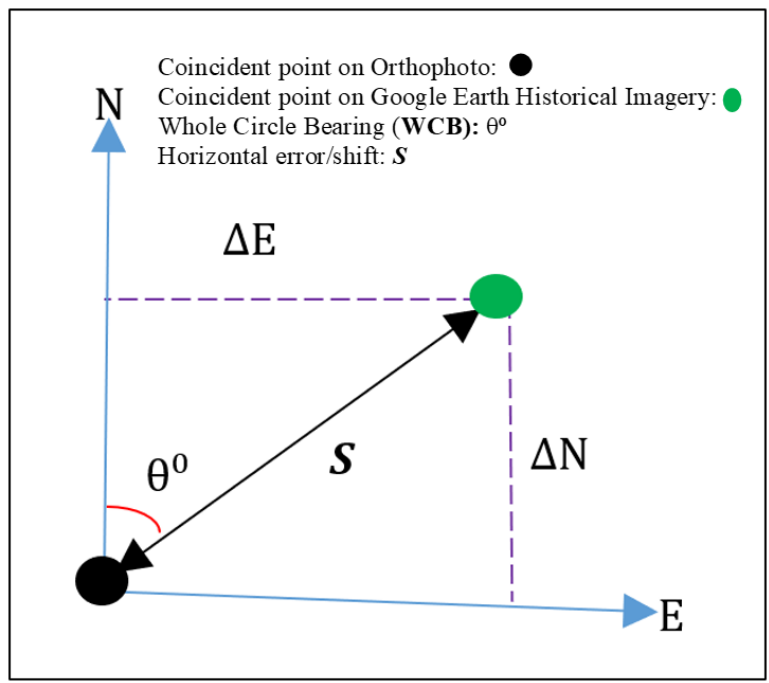

Figure 12: Illustration of horizontal error/shift between a point on the reference orthophoto and the corresponding position on Google Earth

The assessment of the horizontal accuracy is based on the ASPRS Positional Accuracy Standards for Digital Geospatial Data version 1.0 (ASPRS, 2014). Horizontal accuracy was assessed using root mean square error (RMSE) statistics in the horizontal plane (i.e., RMSEx, RMSEy and RMSEr), while the vertical accuracy was assessed in the z-dimension. In addition, the standard deviation (SD) and standard error of the mean (SEM) of the coordinate differences were also calculated. Figure 13 shows an overlay of orthophoto reference points and GE points in two areas within the University of Lagos. ASPRS (2014) defines positional accuracy as "the accuracy of the position of features, including horizontal and vertical positions, with respect to horizontal and vertical datums." Based on the ASPRS recommendations for imagery at an estimated pixel size of 60cm, the Google Earth images used in this study are expected to meet the ASPRS Accuracy Standards of 120cm RMSEx and RMSEy Horizontal Accuracy Class for Standard Mapping and GIS work. The absolute horizontal accuracy should be less than or equal to 2.448 * 120cm at the 95% confidence level. The corresponding estimates of horizontal accuracy at the 95% Confidence level were computed using methodologies documented in the National Standard for Spatial Data Accuracy (NSSDA). The ASPRS standards for vertical accuracy specify absolute vertical accuracy measures for various classes of elevation data based on their spatial resolution (ranging from 1cm to 333.3cm). However, due to the uncertainty surrounding the source of GE elevations including its resolution characteristics, the vertical accuracy assessment is based solely on the reference GCPs. It was computed using the RMSE statistics for non-vegetated terrain (most of the 558 GCPs are located in non-vegetated terrain).

The horizontal RMSE ($RMSE_r$) is given as:

$$RMSE_r \quad = \quad \sqrt{(RMSE_X)^2 + (RMSE_Y)^2}$$

$$= \sqrt{\sum[\Delta E^2 + \Delta N^2]/n} \tag{3}$$

The horizontal linear RMSE in the X direction (Easting) is given below:

$$RMSE_X = \sqrt{\sum (E_{GE} - E_{OP})^2/n} \tag{4}$$

The horizontal linear RMSE in the Y direction (Northing) is given below:

$$RMSE_Y = \sqrt{\sum (N_{GE} - N_{OP})^2/n} \tag{5}$$

Where:

n  - the number of check points

NSSDA Horizontal Accuracy at 95% Confidence Level = RMSEr × 1.7308 $\qquad$ (6)

NSSDA Vertical Accuracy = RMSEz $\qquad$ (7)

$$RMSE_Z = \sqrt{\sum (Z_i - Z_j)^2/n} \tag{8}$$

$\quad$ $Z_i$ = GCP elevation value

$\quad$ $Z_j$ = Google Earth, SRTM or AW3D elevation value.

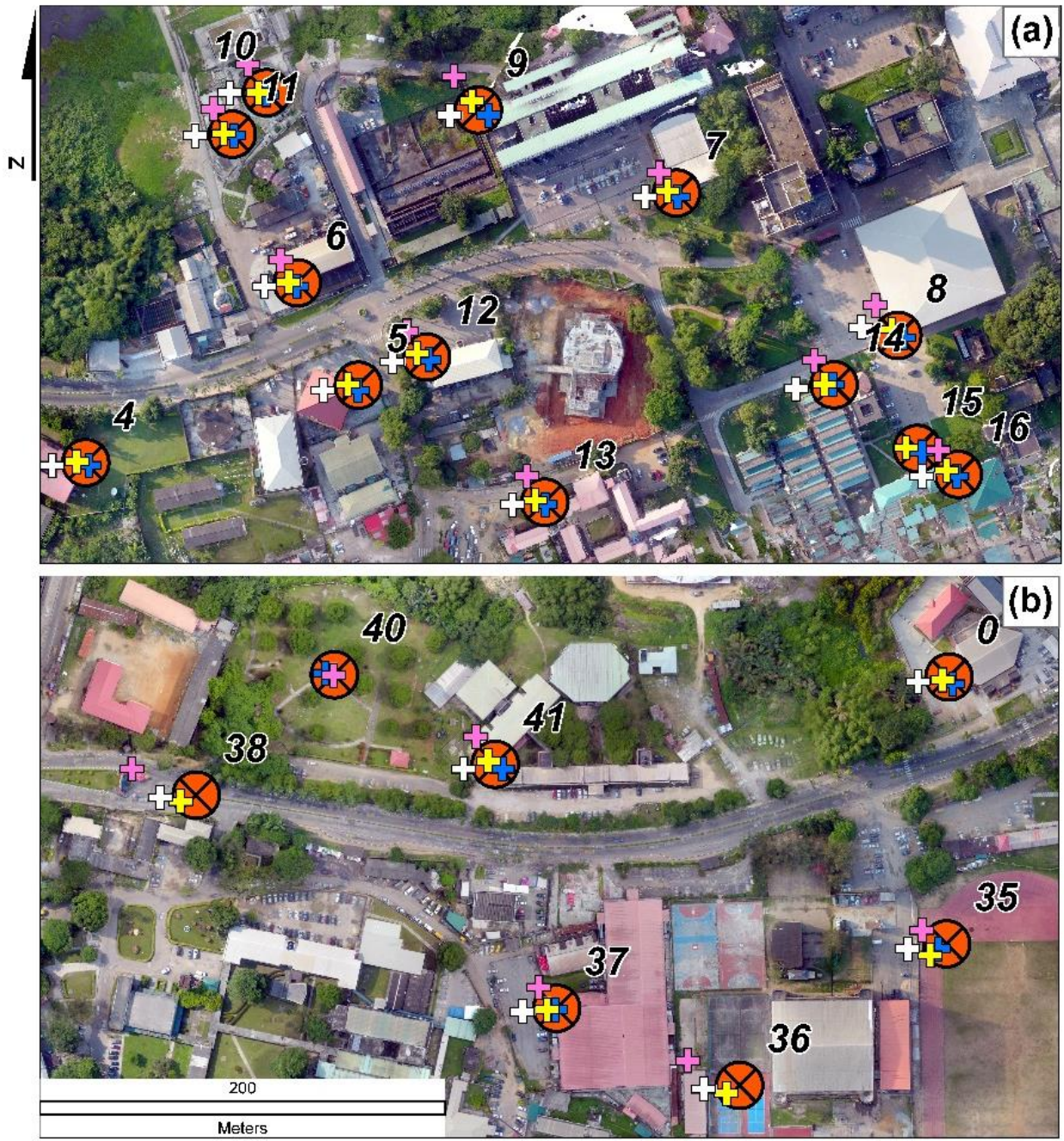

Figure 13: An overlay of orthophoto reference points and GE points in two areas within the University of Lagos, (a) north, and (b) west. The reference points are displayed as red circles, and corresonding GE points are displayed as crosses (2000 - pink, 2008 - white, 2012 – blue and 2018 -yellow)

## 3.    RESULTS AND DISCUSSION
The result of the findings for this study are analysed and discussed for both horizontal and vertical accuracy.

### 3.1 Analysis of Horizontal Accuracy

Table 1 shows the coordinates of the orthophoto reference points and the Google Earth (GE) coincident points. The coordinates are presented in the Universal Transverse Mercator (UTM) system – Eastings and Northings. Table 2 presents the coordinate differences between the reference points and the GE points. The difference in the eastings of the orthophoto reference points and GE coincident points ranged from $0 – 11.99m$ in the year 2000, $2.98 – 18.99m$ in the year 2008, $0.04 – 15.98m$ in the year 2012 and $1.14 – 7.99m$ in the year 2018. In the northings, the differences ranged from $0 – 24.92m$ in the year 2000, $0.11 – 10.32m$ in the year 2008, $0.09 – 10.32m$ in the year 2012 and $0.04 – 11.92m$ in the year 2018. The SDs and RMSEs of the differences in eastings were derived as follows: year 2000 (SD: 4.97m; RMSE$_x$: 9.52m), year 2008 (SD: 4.34m; RMSE$_x$: 15.97m), year 2012 (SD: 4.36m; RMSE$_x$: 4.93m) and year 2018 (SD: 1.48m; RMSE$_x$: 4.39m). The SDs and RMSEs of the differences in northings were derived as follows: year 2000 (SD: 4.83m; RMSE$_y$: 14.05m), year 2008 (SD: 3.71m; RMSE$_y$: 3.72m), year 2012 (SD: 3.54m; RMSE$_y$: 3.65m) and year 2018 (SD: 3.66m; RMSE$_y$: 4.28m).

The differences in eastings and northings of the orthophoto reference points and GE coincident points are represented by stacked columns in Figures 14a and 14b respectively. Consistent $\Delta E$ negative values in the years 2000, 2008, 2018 are suggestive of a systematic error that displaces the Google Earth imagery westwards from the true positions. The largest easting differences are observed in the 2008 imagery with some exceeding 15m while the least differences are observed in the 2012 imagery. The highest horizontal linear RMSE in the x direction (RMSE$_x$) is observed in the 2008 imagery (15.975m), whereas the least horizontal linear RMSE is observed in the 2018 imagery (4.385m). It can be seen that the horizontal differences (Figure 14a) oscillate gently about the mean differences, with spiked differences between points 22 and 27 for the earliest 3 years. Surprisingly, the year 2000 imagery seems to perform better than 2008 in terms of the magnitude of its differences relative to the source of groundtruth (orthophotos). In 2012, the easting coordinates are generally better fitted with absolute differences closer to zero than in prior years. However, it is interesting to note the spikes at some points such as #25 which indicate anomalies in the image planimetry. In 2018, there is a generally better fit for all easting coordinates, with differences very close to zero and the elimination of the spike, which had probably been rectified in this recent version of the imagery.

The differences in northings ($\Delta N$) are shown in Figure 14b. For years 2000 and 2018, the differences are mostly positive and fairly consistent, whereas the differences for years 2008 and 2012 are mostly negative. The trend suggests a systematic or consistent shift in the positions of the GE images generally occurring in the north-south axis; towards the south for years 2000 and 2018, and northwards for years 2008 and 2012. However, the magnitude of the shift is relatively large northwards but minimal southwards. In terms of the horizontal linear error in the y direction RMSE (RMSE$_y$), year 2000 imagery has the highest RMSE of 14.046m while year 2012 imagery has the least RMSE of 3.646m. The northing differences from year 2000 are generally worse off by a significant measure than all other periods. Its absolute differences generally range from 10 to 25m, with the exception of point 30 whose absolute difference is exactly zero. Spikes at points 19 and 20 are seen at the images for 2008 to 2018. Years 2008 and 2012 images generally yield northing differences that are closest to zero.

Table 1: Coordinates of the orthophoto reference points and GE points

| ID | Orthophoto | | GE - 2000 | | GE - 2008 | | GE - 2012 | | GE - 2018 | |
|---|---|---|---|---|---|---|---|---|---|---|
| | Easting (m) | Northing (m) | Easting (m) | Northing (m) | Easting (m) | Northing (m) | Easting (m) | Northing (m) | Easting (m) | Northing (m) |
| 1 | 543316.50 | 720678.04 | 543309 | 720690 | 543301 | 720676 | 543317 | 720676 | 543313 | 720678 |
| 3 | 543527.11 | 720608.96 | 543517 | 720621 | 543510 | 720606 | 543530 | 720606 | 543522 | 720610 |
| 6 | 543859.21 | 720597.77 | 543851 | 720609 | 543843 | 720596 | 543859 | 720596 | 543855 | 720598 |
| 7 | 544045.31 | 720641.96 | 544037 | 720652 | 544030 | 720640 | 544046 | 720640 | 544041 | 720643 |
| 8 | 544153.99 | 720572.17 | 544143 | 720587 | 544136 | 720576 | 544157 | 720571 | 544151 | 720575 |
| 9 | 543947.95 | 720682.63 | 543936 | 720699 | 543934 | 720680 | 543953 | 720680 | 543945 | 720687 |
| 10 | 543844.04 | 720691.11 | 543835 | 720705 | 543826 | 720691 | 543844 | 720690 | 543840 | 720691 |
| 11 | 543827.14 | 720670.66 | 543818 | 720683 | 543809 | 720670 | 543828 | 720670 | 543823 | 720671 |
| 12 | 543922.67 | 720561.41 | 543913 | 720574 | 543906 | 720559 | 543924 | 720560 | 543918 | 720563 |
| 13 | 543980.75 | 720490.21 | 543972 | 720503 | 543965 | 720489 | 543981 | 720489 | 543976 | 720491 |
| 14 | 544122.65 | 720547.91 | 544113 | 720560 | 544104 | 720546 | 544122 | 720548 | 544118 | 720548 |
| 16 | 544183.70 | 720504.00 | 544174 | 720516 | 544166 | 720503 | 544183 | 720503 | 544178 | 720506 |
| 19 | 544162.74 | 720246.68 | 544151 | 720268 | 544144 | 720257 | 544160 | 720257 | 544156 | 720258 |
| 20 | 544125.86 | 720197.08 | 544114 | 720222 | 544109 | 720207 | 544125 | 720206 | 544120 | 720209 |
| 22 | 544004.85 | 720190.64 | 543993 | 720206 | 543988 | 720190 | 544009 | 720189 | 544001 | 720195 |
| 25 | 543897.02 | 719983.64 | 543908 | 719992 | 543900 | 719980 | 543913 | 719978 | 543894 | 719986 |
| 26 | 543925.06 | 719973.52 | 543927 | 719984 | 543917 | 719977 | 543932 | 719976 | 543922 | 719975 |
| 27 | 543243.01 | 719891.69 | 543236 | 719903 | 543225 | 719890 | 543241 | 719890 | 543238 | 719890 |
| 29 | 543241.67 | 719747.52 | 543233 | 719760 | 543226 | 719748 | 543244 | 719748 | 543238 | 719751 |
| 30 | 543244.93 | 719734.02 | 543244 | 719734 | 543229 | 719732 | 543248 | 719731 | 543240 | 719735 |
| 32 | 543214.77 | 719927.15 | 543206 | 719940 | 543198 | 719925 | 543216 | 719924 | 543211 | 719928 |
| 35 | 542994.99 | 720368.45 | 542983 | 720375 | 542976 | 720366 | 542991 | 720366 | 542987 | 720363 |
| 37 | 542803.65 | 720336.67 | 542795 | 720347 | 542787 | 720335 | 542803 | 720336 | 542799 | 720336 |
| 41 | 542773.38 | 720456.37 | 542764 | 720470 | 542758 | 720454 | 542777 | 720454 | 542770 | 720458 |
| 42 | 543884.70 | 719262.76 | 543875 | 719281 | 543872 | 719258 | 543894 | 719258 | 543883 | 719268 |
| 43 | 543869.93 | 719167.54 | 543859 | 719187 | 543856 | 719163 | 543879 | 719163 | 543868 | 719174 |
| 44 | 543859.14 | 719099.52 | 543849 | 719119 | 543846 | 719095 | 543867 | 719095 | 543858 | 719106 |

Table 2: Coordinate differences between the orthophoto reference points and the GE points

| ID | ΔE (m) | | | | ID | ΔN (m) | | | |
|---|---|---|---|---|---|---|---|---|---|
| | Year 2000 | Year 2008 | Year 2012 | Year 2018 | | Year 2000 | Year 2008 | Year 2012 | Year 2018 |
| 1 | -7.50 | -15.50 | 0.50 | -3.50 | 1 | 11.96 | -2.04 | -2.04 | -0.04 |
| 3 | -10.11 | -17.11 | 2.89 | -5.11 | 3 | 12.04 | -2.96 | -2.96 | 1.04 |
| 6 | -8.21 | -16.21 | -0.21 | -4.21 | 6 | 11.23 | -1.77 | -1.77 | 0.23 |
| 7 | -8.31 | -15.31 | 0.69 | -4.31 | 7 | 10.04 | -1.96 | -1.96 | 1.04 |
| 8 | -10.99 | -17.99 | 3.01 | -2.99 | 8 | 14.83 | 3.83 | -1.17 | 2.83 |
| 9 | -11.95 | -13.95 | 5.05 | -2.95 | 9 | 16.37 | -2.63 | -2.63 | 4.37 |
| 10 | -9.04 | -18.04 | -0.04 | -4.04 | 10 | 13.89 | -0.11 | -1.11 | -0.11 |
| 11 | -9.14 | -18.14 | 0.86 | -4.14 | 11 | 12.34 | -0.66 | -0.66 | 0.34 |
| 12 | -9.67 | -16.67 | 1.33 | -4.67 | 12 | 12.59 | -2.41 | -1.41 | 1.59 |
| 13 | -8.75 | -15.75 | 0.25 | -4.75 | 13 | 12.79 | -1.21 | -1.21 | 0.79 |
| 14 | -9.65 | -18.65 | -0.65 | -4.65 | 14 | 12.09 | -1.91 | 0.09 | 0.09 |
| 16 | -9.70 | -17.70 | -0.70 | -5.70 | 16 | 12.00 | -1.00 | -1.00 | 2.00 |
| 19 | -11.74 | -18.74 | -2.74 | -6.74 | 19 | 21.32 | 10.32 | 10.32 | 11.32 |
| 20 | -11.86 | -16.86 | -0.86 | -5.86 | 20 | 24.92 | 9.92 | 8.92 | 11.92 |
| 22 | -11.85 | -16.85 | 4.15 | -3.85 | 22 | 15.36 | -0.64 | -1.64 | 4.36 |
| 25 | 10.98 | 2.98 | 15.98 | -3.02 | 25 | 8.36 | -3.64 | -5.64 | 2.36 |
| 26 | 1.94 | -8.06 | 6.94 | -3.06 | 26 | 10.48 | 3.48 | 2.48 | 1.48 |
| 27 | -7.01 | -18.01 | -2.01 | -5.01 | 27 | 11.31 | -1.69 | -1.69 | -1.69 |
| 29 | -8.67 | -15.67 | 2.33 | -3.67 | 29 | 12.48 | 0.48 | 0.48 | 3.48 |
| 30 | 0.00 | -15.93 | 3.07 | -4.93 | 30 | 0.00 | -2.02 | -3.02 | 0.98 |
| 32 | -8.77 | -16.77 | 1.23 | -3.77 | 32 | 12.85 | -2.15 | -3.15 | 0.85 |
| 35 | -11.99 | -18.99 | -3.99 | -7.99 | 35 | 6.55 | -2.45 | -2.45 | -5.45 |
| 37 | -8.65 | -16.65 | -0.65 | -4.65 | 37 | 10.33 | -1.67 | -0.67 | -0.67 |
| 41 | -9.38 | -15.38 | 3.62 | -3.38 | 41 | 13.63 | -2.37 | -2.37 | 1.63 |
| 42 | -9.70 | -12.70 | 9.30 | -1.70 | 42 | 18.24 | -4.76 | -4.76 | 5.24 |
| 43 | -10.93 | -13.93 | 9.07 | -1.93 | 43 | 19.46 | -4.54 | -4.54 | 6.46 |
| 44 | -10.14 | -13.14 | 7.86 | -1.14 | 44 | 19.48 | -4.52 | -4.52 | 6.48 |
| Mean | -8.18 | -15.40 | 2.45 | -4.14 | 21.56 | 13.22 | -0.78 | -1.11 | 2.33 |
| SD | 4.97 | 4.34 | 4.36 | 1.48 | 13.22 | 4.83 | 3.71 | 3.54 | 3.66 |
| RMSEx | 9.523 | 15.975 | 4.931 | 4.385 | RMSEy | 14.046 | 3.724 | 3.646 | 4.278 |

Figure 14c shows the horizontal errors/shifts in Google Earth images. Table 3 shows the horizontal errors in the GE points including the horizontal linear RMSE in the radial direction that includes both x- and y- coordinate errors (RMSEr). The errors ranged from 0 – 27.6m in year 2000, 4.71 – 21.39m in year 2008, 0.65 – 16.95m in year 2012 and 3.4 – 13.29m in year 2018. The SD and RMSE of the horizontal errors were derived as follows: year 2000 (SD: 5.05m; RMSEr: 16.969m), year 2008 (SD: 3.29m; RMSEr: 16.404m), year 2012 (SD: 4.01m; RMSEr: 6.133m), and year 2018 (SD: 2.54m; RMSEr: 6.127m). It can be seen that the positional accuracy degrades starting from year 2000 onwards, and it is most accurate in year 2018. Year 2000 shows the widest range of errors, with a vastly undulating trend that represents a very poor fit of points, with the highest shift at 27.6m. There is a slight improvement in year 2008 where RMSEr drops to 16.404m from 16.969 in year 2000. Years 2012 and 2018 display the closest fit to the orthophoto. The SDs generally reduce for later epochs for both easting and northing coordinates, with very slight deviations as shown in the increases in SD of easting coordinate differences from 2008 to 2012, and northing coordinate differences from 2012 to 2018. The trend is slightly different for the RMSEs, where there is a significant increase in RMSEx from 2000 to 2008, before generally dropping off. While for RMSEy, it decreases from 2000 till 2012 before picking up again in 2018.

Following equation 6, the NSSDA horizontal accuracies at 95% Confidence Level were calculated as follows – year 2000 (29.369m), year 2008 (28.391m), year 2012 (10.615m) and year 2018 (10.603m). The most recent imagery (year 2018) is the most accurate followed by the year 2012 imagery. However, the difference in accuracy between 2012 and 2018 is negligible. The images for 2000 and 2008 are the worst in terms of horizontal accuracy.

Table 3: Horizontal errors/shifts (S) in the Google Earth images

| ID | Horizontal error/shift | | | |
|---|---|---|---|---|
| | Year 2000 (m) | Year 2008 (m) | Year 2012 (m) | Year 2018 (m) |
| 1 | 14.12 | 15.63 | 2.10 | 3.50 |
| 3 | 15.72 | 17.36 | 4.14 | 5.21 |
| 6 | 13.91 | 16.30 | 1.78 | 4.22 |
| 7 | 13.03 | 15.44 | 2.08 | 4.43 |
| 8 | 18.46 | 18.39 | 3.23 | 4.12 |
| 9 | 20.27 | 14.19 | 5.70 | 5.27 |
| 10 | 16.57 | 18.04 | 1.11 | 4.04 |
| 11 | 15.36 | 18.16 | 1.08 | 4.16 |
| 12 | 15.87 | 16.84 | 1.94 | 4.93 |
| 13 | 15.50 | 15.80 | 1.23 | 4.82 |
| 14 | 15.47 | 18.75 | 0.65 | 4.65 |
| 16 | 15.43 | 17.73 | 1.22 | 6.04 |
| 19 | 24.33 | 21.39 | 10.67 | 13.17 |
| 20 | 27.60 | 19.57 | 8.96 | 13.29 |
| 22 | 19.40 | 16.87 | 4.46 | 5.82 |
| 25 | 13.80 | 4.71 | 16.95 | 3.83 |
| 26 | 10.66 | 8.78 | 7.37 | 3.40 |
| 27 | 13.30 | 18.09 | 2.63 | 5.29 |
| 29 | 15.20 | 15.68 | 2.38 | 5.06 |
| 30 | 0.00 | 16.05 | 4.31 | 5.02 |
| 32 | 15.56 | 16.91 | 3.38 | 3.87 |
| 35 | 13.67 | 19.15 | 4.68 | 9.67 |
| 37 | 13.47 | 16.73 | 0.93 | 4.70 |
| 41 | 16.54 | 15.56 | 4.33 | 3.75 |
| 42 | 20.65 | 13.56 | 10.45 | 5.50 |
| 43 | 22.32 | 14.66 | 10.14 | 6.74 |
| 44 | 21.96 | 13.89 | 9.07 | 6.57 |
| $RMSE_r$ | 16.969 | 16.404 | 6.133 | 6.127 |
| $RMSE_r$ (95% C.L) | 29.369 | 28.391 | 10.615 | 10.603 |

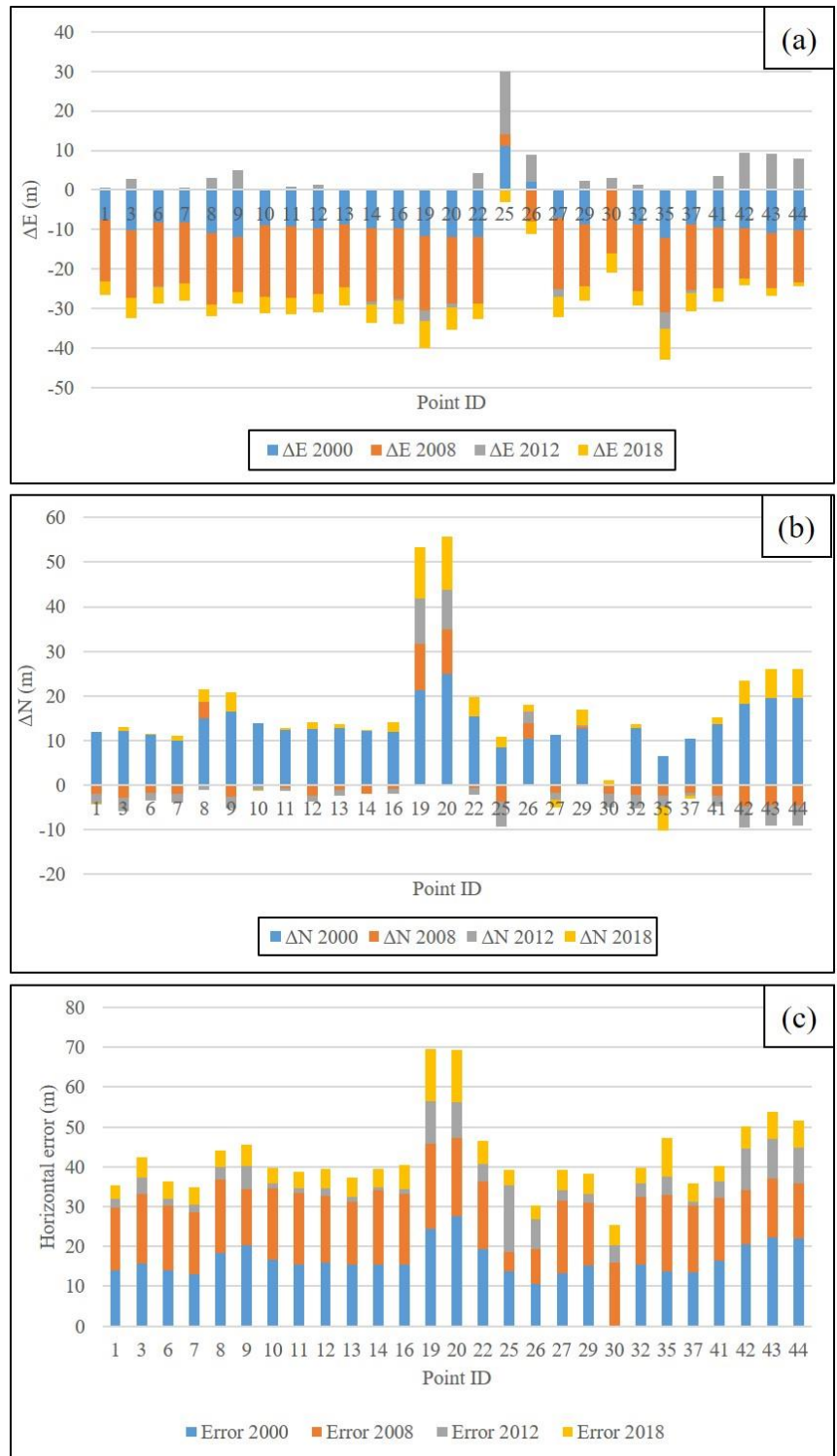

Figure 14: Stacked columns showing (a) differences in eastings of the orthophoto reference points and GE coincident points (b) differences in northings of the orthophoto reference points and GE coincident points, and (c) horizontal errors/shifts in Google Earth images

**3.2     Magnitude and Direction of Horizontal Error**

Table 4 shows the directions and whole circle bearings (WCBs) of the horizontal errors/shifts. In general, the maximum shifts in the four periods of study are in the NW direction: 2000 (Mag: 27.60m; Dir: 334.55º), 2008 (Mag: 21.39m; Dir: 300.47º), 2012 (Mag: 16.95m; Dir: 354.49º) and 2018 (Mag: 13.28m; Dir: 350.02º). The mean quadrants of the shifts are shown in Table 5. With the exception of year 2012, the mean shifts are generally skewed towards the western direction: 2000 (Mag: 16.26m; Dir: 300.70º), 2008 (Mag: 16.08m; Dir: 262.59º), 2018 (Mag: 5.60m; Dir: 294.88º). In the 2000 imagery, the general shift is in the northern direction, more notably NW with mean and maximum displacement of 16.26m and 27.60 respectively in that direction (see Table 5). The mean shifts for the 2000 imagery (16.264m)  and 2008 imagery (16.082m) are within a similar range but there is a sharp drop to 4.703m in year 2012 and in 2018 it rises to 12.142m. Although considerable similarity exists between the mean and maximum shifts in 2000 and 2008, their directions are opposite. For year 2012, the mean shift is 4.70m in the SE direction. A remarkable feature of the 2018 imagery is a general shift that is skewed to the western axis. This is indicated by a mean value of the eastern displacement component (– 2.909m) and a range of values between -1.140m and -7.990m along the eastern/western axis. It is further corroborated by the quadrant/angle of the minimum, maximum and mean shifts. There is a general decrease in the magnitude of the mean shifts from the earliest year to the latest year, with 2018 having the least horizontal error along that axis.

Ninety percent of the displacement in year 2000 occurred around the NW quadrant (Table 4), no displacement exceeds ±13m along the EW axis, and displacement along the NS axis occurred only in the northern quadrant (Figure 15). In the year 2008, about 80% of the displacements occurred in the SW quadrant, with a significant amount of shift occurring within -10m to -20m range on the EW axis.  There is however little error between –10m to +5m as one moves eastward along the EW axis. This may be a strong indication of systematic error that skews position westward away from their true value. The case is slightly different when one considers displacement along the NS axis. Displacement/shift along this axis generally occurs within ±5m from the origin. In the year 2012, the shifts are generally skewed towards the SE quadrant. Along the NS axis displacement generally occur within ±5m. A similar feature is also noticeable in the shift along the EW axis. In the year 2018, the shifts generally occurred towards the NW quadrant. Interestingly no displacement occurred as one moves eastward. Only at three locations were displacement within -2m to -1m observed. A significant number occurred within -6m and -1m along the EW axis. Along the NS axis, displacement majorly occurs within -2m to +6m.

Table 4: Directions and Whole Circle Bearings (WCBs) of the horizontal errors/shifts

| ID | 2000 | | 2008 | | 2012 | | 2018 | |
|---|---|---|---|---|---|---|---|---|
| | **Dir (°)** | **WCB (°)** | **Dir (°)** | **WCB (°)** | **Dir (°)** | **WCB (°)** | **Dir (°)** | **WCB (°)** |
| 1 | -32.09 | 327.91 | 82.50 | 262.50 | -13.77 | 166.23 | 89.35 | 269.35 |
| 3 | -40.02 | 319.98 | 80.19 | 260.19 | -44.31 | 135.69 | -78.50 | 281.50 |
| 6 | -36.17 | 323.83 | 83.77 | 263.77 | 6.77 | 186.77 | -86.87 | 273.13 |
| 7 | -39.61 | 320.39 | 82.70 | 262.70 | -19.39 | 160.61 | -76.43 | 283.57 |
| 8 | -36.54 | 323.46 | -77.98 | 282.02 | -68.76 | 111.24 | -46.57 | 313.43 |
| 9 | -36.13 | 323.87 | 79.32 | 259.32 | -62.49 | 117.51 | -34.02 | 325.98 |
| 10 | -33.06 | 326.94 | 89.65 | 269.65 | 2.06 | 182.06 | 88.44 | 268.44 |
| 11 | -36.53 | 323.47 | 87.92 | 267.92 | -52.50 | 127.50 | -85.31 | 274.69 |
| 12 | -37.53 | 322.47 | 81.77 | 261.77 | -43.33 | 136.67 | -71.20 | 288.80 |
| 13 | -34.38 | 325.62 | 85.61 | 265.61 | -11.67 | 168.33 | -80.56 | 279.44 |
| 14 | -38.60 | 321.40 | 84.15 | 264.15 | -82.12 | 277.88 | -88.89 | 271.11 |
| 16 | -38.95 | 321.05 | 86.77 | 266.77 | 34.99 | 214.99 | -70.67 | 289.33 |
| 19 | -28.84 | 331.16 | -61.16 | 298.84 | -14.87 | 345.13 | -30.77 | 329.23 |
| 20 | -25.45 | 334.55 | -59.53 | 300.47 | -5.51 | 354.49 | -26.18 | 333.82 |
| 22 | -37.65 | 322.35 | 87.83 | 267.83 | -68.44 | 111.56 | -41.45 | 318.55 |
| 25 | 52.72 | 52.72 | -39.31 | 140.69 | -70.56 | 109.44 | -51.99 | 308.01 |
| 26 | 10.49 | 10.49 | -66.65 | 293.35 | 70.34 | 70.34 | -64.19 | 295.81 |
| 27 | -31.79 | 328.21 | 84.64 | 264.64 | 49.94 | 229.94 | 71.36 | 251.36 |
| 29 | -34.79 | 325.21 | -88.25 | 271.75 | 78.36 | 78.36 | -46.52 | 313.48 |
| 30 | 88.77 | 268.77 | 82.77 | 262.77 | -45.47 | 134.53 | -78.76 | 281.24 |
| 32 | -34.31 | 325.69 | 82.69 | 262.69 | -21.33 | 158.67 | -77.29 | 282.71 |
| 35 | -61.35 | 298.65 | 82.65 | 262.65 | 58.45 | 238.45 | 55.70 | 235.70 |
| 37 | -39.94 | 320.06 | 84.27 | 264.27 | 44.13 | 224.13 | 81.80 | 261.80 |
| 41 | -34.54 | 325.46 | 81.24 | 261.24 | -56.79 | 123.21 | -64.25 | 295.75 |
| 42 | -28.00 | 332.00 | 69.45 | 249.45 | -62.90 | 117.10 | -17.97 | 342.03 |
| 43 | -29.32 | 330.68 | 71.95 | 251.95 | -63.41 | 116.59 | -16.63 | 343.37 |
| 44 | -27.50 | 332.50 | 71.02 | 251.02 | -60.10 | 119.90 | -9.98 | 350.02 |

Table 5: Summary statistics of the magnitudes and directions of horizontal error

| Year | Shift | ΔE (m) | ΔN (m) | Magnitude (m) | Direction (°) | Quadrant |
|------|-------|--------|--------|---------------|---------------|----------|
| 2000 | Max. | 10.98 | 24.92 | 27.598 | 334.549 | NW |
|      | Mean | -8.212 | 13.219 | 16.264 | 300.699 | NW |
|      | Min. | -11.990 | -0.020 | 0.930 | 10.488 | NE |
| 2008 | Max. | 2.980 | 10.320 | 21.394 | 300.471 | NW |
|      | Mean | -15.397 | -0.781 | 16.082 | 262.593 | SW |
|      | Min. | -18.990 | -4.760 | 4.704 | 140.693 | SE |
| 2012 | Max. | 15.980 | 10.320 | 16.946 | 354.493 | NW |
|      | Mean | 2.455 | -1.114 | 4.703 | 167.309 | SE |
|      | Min. | -3.990 | -5.640 | 0.656 | 70.336 | NE |
| 2018 | Max. | -1.140 | 11.920 | 13.283 | 350.022 | NW |
|      | Mean | -2.909 | 5.163 | 12.142 | 294.875 | NW |
|      | Min. | -7.990 | -5.450 | 3.399 | 235.702 | SW |

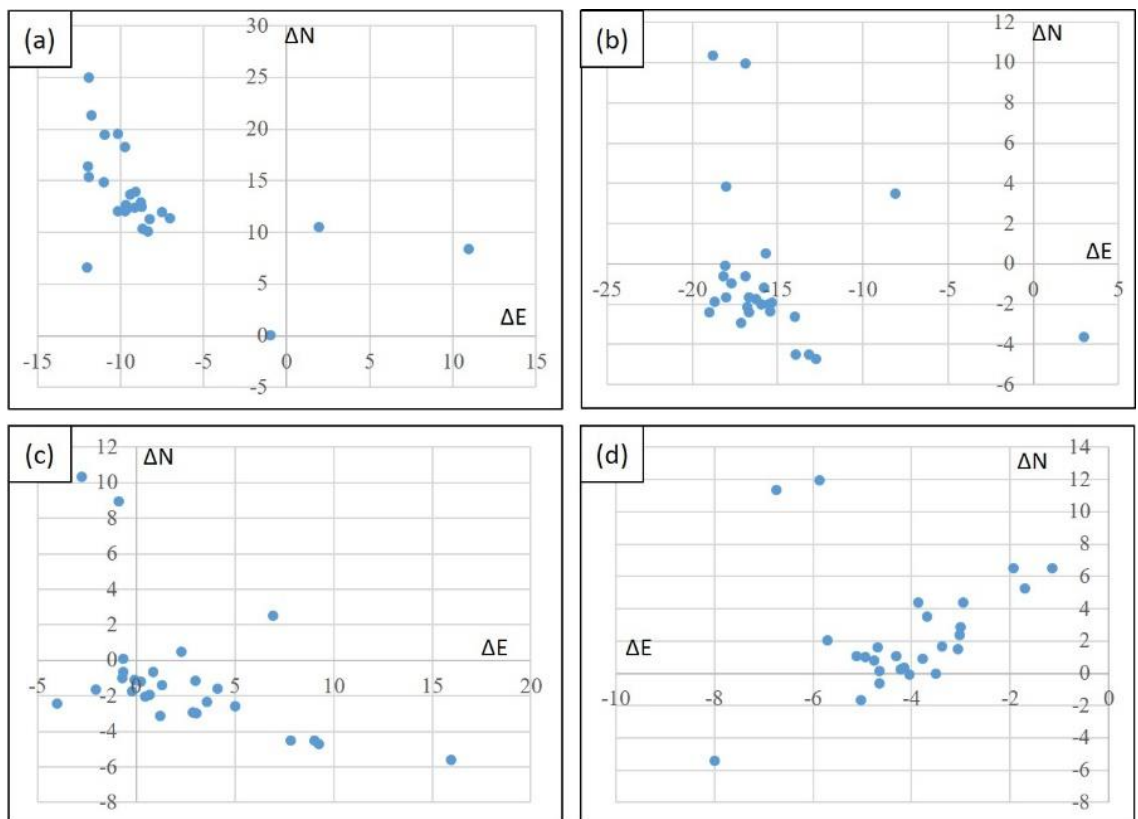

Figure 15: Magnitudes and directions of horizontal error in the Google earth historical images - (a) year 2000 (b) year 2008 (c) year 2012 and (d) year 2018

### 3.3 Analysis of Vertical Accuracy

This section presents the analysis of the vertical accuracy of GE heights ($H_{GE}$), including the comparison with the AW3D-30 v2.1 ($H_{AW3D}$) and SRTM v3.0 ($H_{SRTM}$) DEMs. The basis of the assessment was the heights (elevations) of 558 GCPs spread around Lagos State. The heights from all three datasets fell within the general range of $0 - 73$m. The mean heights are as follows: GCP (16.410m), Google Earth (16.588m) AW3D (18.482m) and SRTM (18.188m). Table 6 presents the SDs, SEMs and RMSEs of the height differences. The lowest RMSE of 3.682m and highest vertical accuracy was observed in SRTM followed by AW3D with an RMSE of 4.388m. The highest RMSE of 6.213m and lowest vertical accuracy occurred in Google Earth. The mean height differences in GE are the least at 0.178m. However, the range of GE elevations is the highest (64.152m) and this shows that the elevations have a wider spread relative to the GCPs. These results show the low reliability of estimated elevations from GE relative to AW3D and SRTM DEM. This conclusion is also supported by its high values for RMSE and SD, indicating a much larger spread of elevations about the mean, relative to other DEMs in comparison.

Statistics from the SRTM DEM suggest it as the most reliable of all three height estimation products, as it has the least range of height differences, a relatively modest mean elevation error and the least standard deviation, which implies a low spread of elevations about its mean value. Hence, we can safely conclude it is a more reliable product for estimating heights of objects or points than AW3D and GE historical imagery. Finally, AW3D has a slightly lower elevation range than GE, the highest mean elevation and an SD slightly higher than that of SRTM. GE elevations are the least reliable while SRTM is the most reliable.

Table 6: SD and RMSE of the height differences

|  | Range | Minimum | Maximum | Mean | SEM | SD | $RMSE_z$ |
|---|---|---|---|---|---|---|---|
| $\Delta H_{GE-GCP}$ | 64.152 | -28.460 | 35.692 | 0.178 | 0.263 | 6.216 | 6.213 |
| $\Delta H_{AW3D-GCP}$ | 58.850 | -15.981 | 42.869 | 2.072 | 0.164 | 3.872 | 4.388 |
| $\Delta H_{SRTM-GCP}$ | 35.569 | -22.981 | 12.587 | 1.778 | 0.137 | 3.228 | 3.682 |

Figure 16 shows histograms of height differences in the three datasets. SRTM DEMs have the least spread of height differences, with significant height differences ranging from just less than -10m to over 10m. Also, the relative frequency of various height differences has a near-normal distribution, suggesting reliability and predictability. The most commonly occurring height differences are near the zero mark, suggesting a mean close to zero. The GE histogram has a similar bell shape for its height differences. The most commonly occurring differences and mean also hover about the zero point, with nearly evenly distributed differences, although slightly skewed to the left. Height differences for AW3D are significantly skewed to the right, with its most commonly occurring elevation differences close to zero. However, its obvious range for differences is much lower than that of GE which is by far the greatest and probably the least reliable.

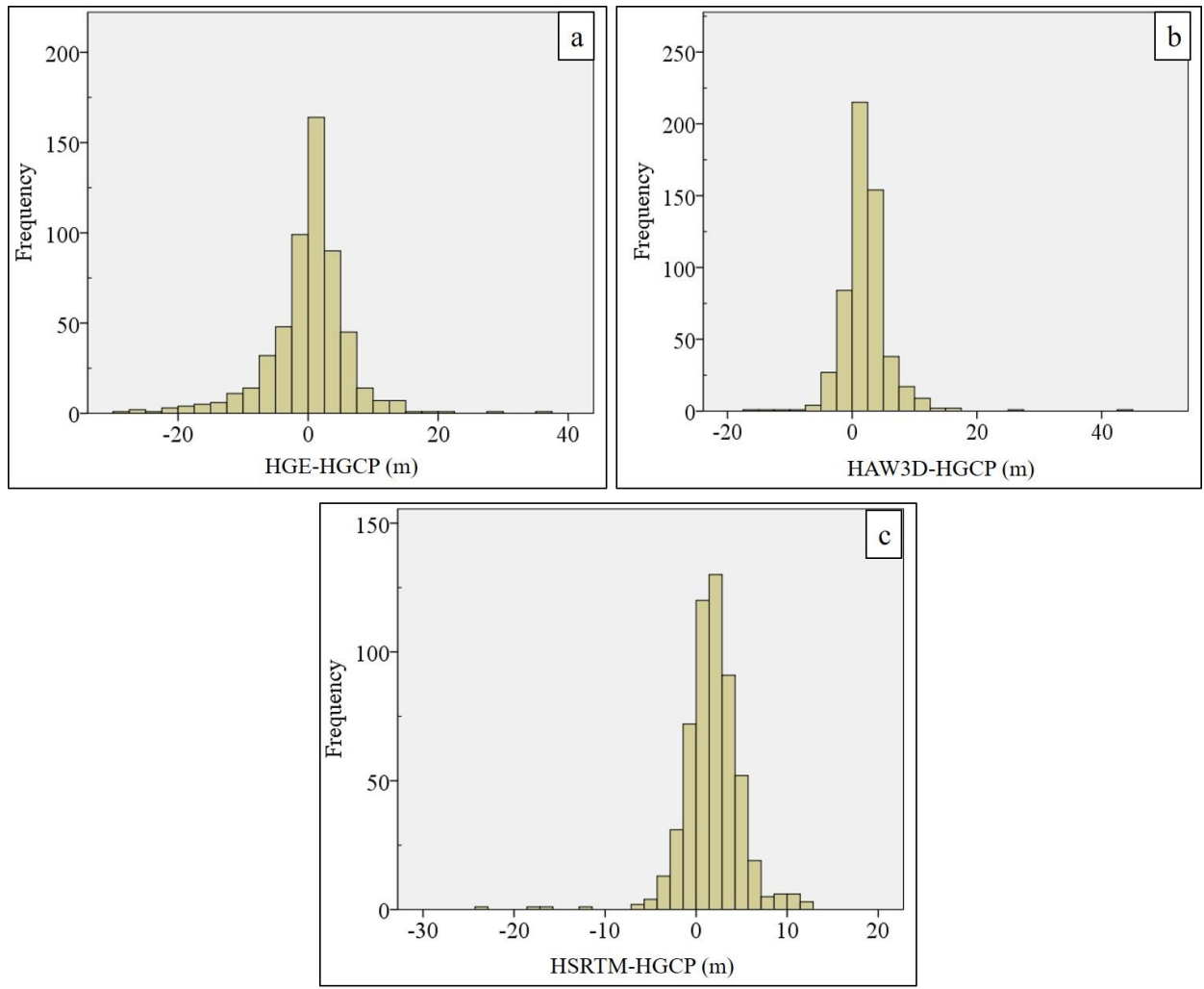

Figure 16: Histogram of height differences (a) Google Earth elevations (b) AW3D (c) SRTM

## 4. DISCUSSION

The difference between 2018 and 2012 images are quite negligible, hence they can be said to be of similar accuracy. The images for 2000 and 2008 are the worst in terms of horizontal accuracy. The accuracy is lowest at year 2000 and gradually improves over time till 2018. This suggests that Google has made efforts at improving the quality of its satellite imagery over the years. Also, it shows there is a continuous enhancement in the accuracy and reliability of satellite imagery data sources which form the source of Google Earth data. In terms of the vertical accuracy, Google Earth elevation data had the highest RMSE of 6.213m followed by AW3D with an RMSE of 4.388m and SRTM with an RMSE of 3.682m. A crucial feature worthy of note are the tendency of significant variability in accuracy of GE images within a relatively small area of study. The noticeable spikes at certain points within the area of study (e.g. 19 and 20) supports this fact. It would therefore be safe to note that accuracy may vary greatly between points within the same study area. Another fact that becomes clear from this study is thus: GE images are significantly skewed westward within the study area. With the exception of the 2012 imagery which may be regarded as having a mean SW skewness, all other images tend to be skewed towards the NW direction. This skewness towards certain direction which strongly indicates the presence of a systematic error, must be taken into account when using the imagery. More research is needed in other locations to determine if this skewness is a general trend with GE imagery.

Although the vertical accuracy of SRTM and AW3D are superior, Google Earth still presents clear advantages in terms of its ease-of-use and contextual awareness. For example, a wide variety of end-users and software developers including those not allied to spatial sciences are able to easily interact with GE data and utilise it for their applications. Google Earth APIs (application programming interfaces) enable developers to embed Google Earth imagery into web applications. Moreover, Google Earth is ubiquitous, and it has a very user-friendly interface that appeals to users from all spheres of life. SRTM and AW3D are distributed via online data archives in which the processes for search and retrieval are not straightforward for citizens without some background knowledge or exposure to spatial information sciences.

## 5.     CONCLUSION AND RECOMMENDATIONS

This study assessed the positional (horizontal and vertical) accuracy of Google Earth historical imagery in Lagos Nigeria. The horizontal accuracy was assessed in terms of the RMSE using historical images of the University of Lagos acquired at four epochs – 2000, 2008, 2012 and 2018. In accordance with the ASPRS recommendations for images at an estimated pixel size of 60cm, the Google Earth images assessed in this study are expected to meet the ASPRS Accuracy Standards of 120cm RMSEx and RMSEy Horizontal Accuracy Class for Standard Mapping and GIS work. Also, the absolute horizontal accuracy should be less than or equal to 2.448 * 120cm at the 95% confidence level. The horizontal accuracies were estimated as follows – year 2000 (29.369m), year 2008 (28.391m), year 2012 (10.615m) and year 2018 (10.603m). It is evident that the absolute horizontal accuracies of the GE historical images in the four years considered fell short of the ASPRS standard. In terms of the vertical accuracy, Google Earth elevation data had the highest RMSE of 6.213m and thus the lowest accuracy. The SRTM DEM had the highest vertical accuracy with an RMSE of 3.682m followed by AW3D with an RMSE of 4.388m. These findings strenghten the chain of evidence on the positional accuracies of Google Earth imagery in previous sudies (e.g. Ubukawa, 2013 and Paredes-Hernández et al., 2013).

The main contribution of this study lies in the relevance of the findings to end-users of Google Earth images and to the broad Digital Earth community. It presents an informed perspective through quantitative analysis, on the limitations of historical archive of Google Earth's  imagery in terms of the horizontal accuracy and also, the vertical accuracy of the elevation data. It is clear from literature that variations exist in map and satellite information thereby making the services rendered as error-prone. Users of these services are advised to use GE map and satellite information with caution as stated in GE terms of service. It is also important to point out that the positional accuracy of GE can be enhanced by carrying out image geometric registration on the imagery using established geodetic control points. By carrying out this enhancement, the application of GE imagery becomes extensive. The trends for GE imagery in this study have shown improvement in positional accuracy in more recent imagery. It might be advisable for users to go for the most recent imagery as the trend suggests higher accuracy in recent times. Also, users should not rely on just GE imagery but instead acquire supporting data for their analysis. It is reasonable to augment GE acquired data with other relevant and more accurate sources.

## ACKNOWLEDGEMENTS

We thank Google LLC for free access to Goolge Earth imagery used in this research, the Japan Aerospace Exploration Agency (JAXA) for free access to AW3D-30m DEM and the National Aeronautics and Space Administration (NASA)/United States Geological Surveys (USGS) for free access to SRTM data. The University of Lagos management is also appreciated for permitting us to conduct the UAV survey within the campus. We are also grateful to the Lagos State Surveyor General's Office for provision of ground control points. The DJI Phantom 4 Professional UAV and accessories were made available by Geospatial Research Limited Nigeria. The following individuals who assisted with the GPS observation, UAV survey and orthophoto processing are acknowledged – Kayode Omolaye and Olumide Awe ( Geospatial Research Limited Nigeria), Adefemi Alabi Geosys Nigeria Limited, Shittu Ibrahim (Federal School of Surveying, Nigeria), Tochi Nwaoru, Oluwaseyi Isaac and Adepo Rahmatullahi (Department of Surveying and Geoinformatics, University of Lagos, Nigeria).

## FUNDING

This research did not receive any specific funding from the public, private or not-for-profit sectors.

## CONFLICTS OF INTEREST

There is no conflict of interest regarding the publication of this research.